\documentclass[english]{article}
\usepackage[T1]{fontenc}
\usepackage[latin9]{inputenc}
\usepackage{color}
\usepackage{babel}
\usepackage{mathrsfs}
\usepackage{amsmath}
\usepackage{amssymb}
\usepackage{graphicx}
\usepackage{setspace}
\usepackage[unicode=true,pdfusetitle,
 bookmarks=true,bookmarksnumbered=false,bookmarksopen=false,
 breaklinks=false,pdfborder={0 0 1},backref=false,colorlinks=false]
 {hyperref}

\makeatletter

\providecommand{\tabularnewline}{\\}

\usepackage{cite}

\pdfoutput=1

\usepackage{graphicx}
\DeclareMathOperator*{\name}{\raisebox{-0.8ex}{\scalebox{2.5}{$\ast$}}}

\makeatother

\begin{document}
\noindent \begin{flushleft}
\textbf{\Large{}Stationary-State Statistics of a Binary Neural Network
Model with Quenched Disorder}\\
\par\end{flushleft}{\Large \par}

\noindent \begin{flushleft}
Diego Fasoli$^{1,\ast}$, Stefano Panzeri$^{1}$
\par\end{flushleft}

\medskip{}
\noindent \begin{flushleft}
\textbf{{1} Laboratory of Neural Computation, Center for Neuroscience
and Cognitive Systems @UniTn, Istituto Italiano di Tecnologia, 38068
Rovereto, Italy}
\par\end{flushleft}

\noindent \begin{flushleft}
\textbf{$\ast$ Corresponding Author. E-mail: diego.fasoli@iit.it}
\par\end{flushleft}

\section*{Abstract}

We study the statistical properties of the stationary firing-rate
states of a neural network model with quenched disorder. The model
has arbitrary size, discrete-time evolution equations and binary firing
rates, while the topology and the strength of the synaptic connections
are randomly generated from known, generally arbitrary, probability
distributions. We derived semi-analytical expressions of the occurrence
probability of the stationary states and the mean multistability diagram
of the model, in terms of the distribution of the synaptic connections
and of the external stimuli to the network. Our calculations rely
on the probability distribution of the bifurcation points of the stationary
states with respect to the external stimuli, which can be calculated
in terms of the permanent of special matrices, according to extreme
value theory. While our semi-analytical expressions are exact for
any size of the network and for any distribution of the synaptic connections,
we also specialized our calculations to the case of statistically-homogeneous
multi-population networks. In the specific case of this network topology,
we calculated analytically the permanent, obtaining a compact formula
that outperforms of several orders of magnitude the Balasubramanian-Bax-Franklin-Glynn
algorithm. To conclude, by applying the Fisher-Tippett-Gnedenko theorem,
we derived asymptotic expressions of the stationary-state statistics
of multi-population networks in the large-network-size limit, in terms
of the Gumbel (double exponential) distribution. We also provide a
Python implementation of our formulas and some examples of the results
generated by the code.\\

\noindent \textbf{Keywords:} stationary states, multistability, binary
neural network, quenched disorder, bifurcations, extreme value theory,
order statistics, matrix permanent, Fisher-Tippett-Gnedenko theorem,
Gumbel distribution

\section{Introduction \label{sec:Introduction}}

Biological networks are typically characterized by highly heterogeneous
synaptic connections. Variability in the biophysical properties of
the synapses has been observed experimentally between different types
of synapses, as well as within a given synaptic type \cite{Marder2006}.
Among these properties, the synaptic weights represent the dynamically
regulated strength of connections between pairs of neurons, which
quantify the influence the firing of one neuron has on another. The
magnitude of the synaptic weights is determined by several variable
factors, which in the case, for example, of chemical synapses, include
the size of the available vesicle pool, the release probability of
neurotransmitters into the synaptic cleft, the amount of neurotransmitter
that can be absorbed in the postsynaptic neuron, the efficiency of
neurotransmitter replenishment, etc \cite{Dobrunz1997,Waters2002,Parker2003,Marder2006,Branco2009}.
These factors differ from synapse to synapse, causing heterogeneity
in the magnitude of the synaptic weights. An additional source of
variability in biological networks is represented by the number of
connections made by the axon terminals to the dendrites of post-synaptic
neurons, which affects both the magnitude of the synaptic weights
and the local topology of neural circuits \cite{Branco2009}. Heterogeneity
is an intrinsic network property \cite{Parker2003}, which is likely
to cover an important functional role in preventing neurological disorders
\cite{Santhakumar2004}.

Biologically realistic neural network models should capture most of
the common features of real networks. In particular, synaptic heterogeneity
is typically described by assuming that the synaptic connections can
be modeled by random variables with some known distribution. Under
this assumption, the standard deviation of the distribution can be
interpreted as a measure of synaptic heterogeneity. So far, only a
few papers have investigated the dynamical properties of neural networks
with random synaptic connections, see e.g. \cite{Sompolinsky1988,Cessac1995,Faugeras2009,Hermann2012,Cabana2013}.
These papers focused on the special case of fully-connected networks
of graded-rate neurons in the thermodynamic limit. The neurons in
these models are all-to-all connected with unit probability, namely
the network topology is deterministic, while the strength of the synaptic
connections is normally distributed. The synaptic weights are described
by frozen random variables which do not evolve over time, therefore
these models are said to present \textit{quenched disorder}.

In condensed matter physics, quenched disorder describes mathematically
the highly irregular atomic bond structure of amorphous solids known
as \textit{spin glasses} \cite{Sherrington1976,Kirkpatrick1978,DeAlmeida1978,Mezard1984,Mezard1986,Sherrington1999}.
Spin models are characterized by discrete output, therefore the powerful
statistical approaches developed for studying spin glasses can be
adapted for the investigation of networks of binary-rate neurons in
the thermodynamic limit. Typically, the synaptic weights in these
models are chosen according to the Hebbian rule, so that the systems
behave as attractor neural networks, that can be designed for storing
and retrieving some desired patterns of neural activity (see \cite{Coolen1993}
and references therein).

In order to obtain statistically representative and therefore physically
relevant results in mathematical models with quenched disorder, one
needs to average physical observables over the variability of the
connections between the network units \cite{Sherrington1999}. In
other words, one starts by generating several copies or repetitions
of the network, where the weights and/or the topology of the connections
are generated randomly from given probability distributions. Each
copy owns a different set of frozen connections, which affect the
value of physical observables. Therefore physical observables present
a probability distribution over the set of connections. Since measurements
of macroscopic physical observables are dominated by their mean values
\cite{Sherrington1999}, one typically deals with the difficult problem
of calculating averages over the distribution of the connections.

In \cite{Fasoli2018b} we introduced optimized algorithms for investigating
changes in the long-time dynamics of arbitrary-size recurrent networks,
composed of binary-rate neurons that evolve in discrete time steps.
These changes of dynamics are known in mathematical terms as \textit{bifurcations}
\cite{Kuznetsov1998}. In particular, in \cite{Fasoli2018b} we studied
changes in the number of stationary network states and the formation
of neural oscillations, elicited by variations in the external stimuli
to the network. The synaptic weights and the topology of the network
were arbitrary (possibly random and asymmetric), and we studied the
bifurcation structure of single network realizations without averaging
it over several copies of the model.

In the present paper we extended the mathematical formalism and the
algorithms introduced in \cite{Fasoli2018b}, by deriving a complete
semi-analytical description of the statistical properties of the long-time
network states and of the corresponding bifurcations, across network
realizations. For simplicity, we focused on the mathematical characterization
of the stationary states of random networks, while the more difficult
case of neural oscillations will be discussed briefly at the end of
the paper. Unlike spin-glass theory \cite{Sherrington1976,Kirkpatrick1978,DeAlmeida1978,Mezard1984,Mezard1986,Sherrington1999},
we extended our work beyond the calculation of the expected values,
by deriving complete probability distributions. Moreover, unlike previous
work on random networks of graded neurons \cite{Sompolinsky1988,Cessac1995,Faugeras2009,Hermann2012,Cabana2013},
which focused on fully-connected models with normally distributed
weights in the thermodynamic limit, we investigated the more complicated
case of arbitrary-size networks, with arbitrary distribution of the
synaptic weights and of the topology of the connections. Moreover,
unlike spin-glass models of attractor neural networks \cite{Coolen1993},
we did not consider specifically the problem of storing/retrieving
some desired sequences of neural activity patterns. Rather, we determined
the fixed-point attractors, namely the stationary solutions of the
network equations, that are generated by some given arbitrary distribution
of the synaptic connections (i.e. by connections that are not necessarily
designed to store and retrieve some desired patterns).

The range of network architectures that can be studied with our formalism
is wide, and includes networks whose size, number of neural populations,
synaptic weights distribution, topology and sparseness of the connections
are arbitrary. For this reason, performing a detailed analysis of
how all these factors affect the statistical properties of the stationary
states and of their bifurcation points, is not feasible. Rather, the
purpose of this paper is to introduce the mathematical formalism that
allows the statistical analysis of these networks in the long-time
regime, and to show the results it produces when applied to some examples
of neural network architectures.\\

\noindent \textbf{Paper Outline:} In Sec.\textcolor{blue}{~}(\ref{sec:Materials-and-Methods})
we introduced the binary neural network model (SubSec.\textcolor{blue}{~}(\ref{subsec:The-Network-Model}))
and we studied semi-analytically the statistical properties of the
stationary states and of their bifurcation points, provided the probability
distribution of the synaptic connections is known (SubSec.\textcolor{blue}{~}(\ref{subsec:Statistical-Properties-of-the-Network-Model})).
In particular, in SubSec.\textcolor{blue}{~}(\ref{subsec:Probability-Distribution-of-the-Bifurcation-Points})
we calculated the probability distribution of the bifurcation points
in the stimuli space in terms of the permanent of special matrices,
while in SubSec.\textcolor{blue}{~}(\ref{subsec:Mean-Multistability-Diagram})
we used this result to derive the mean multistability diagram of the
model. In SubSecs.\textcolor{blue}{~}(\ref{subsec:Occurrence-Probability-of-the-Stationary-States-for-a-Given-Combination-of-Stimuli})
and (\ref{subsec:Occurrence-Probability-of-the-Stationary-States-Regardless-of-the-Stimuli}),
we calculated the probability that a given firing-rate state is stationary,
for a fixed combination of stimuli and regardless of the value of
the stimuli, respectively. In SubSec.\textcolor{blue}{~}(\ref{subsec:The-Special-Case-of-Multi-Population-Networks-Composed-of-Statistically-Homogeneous-Populations})
we specialized to the case of statistically-homogeneous multi-population
networks of arbitrary size, and we derived a compact expression of
the matrix permanent, while in SubSec.\textcolor{blue}{~}(\ref{subsec:Large-Network-Limit})
we calculated the asymptotic form of the stationary-state statistics
in the large-size limit. In SubSec.\textcolor{blue}{~}(\ref{subsec:Numerical-Simulations})
we described the Monte Carlo techniques that we used for estimating
numerically the statistical properties of the stationary states and
of their bifurcation points. In Sec.~(\ref{sec:Results}) we used
these numerical results to further validate our semi-analytical formulas,
by studying specific examples of network topologies. In Sec\textcolor{blue}{.~}(\ref{sec:Discussion})
we discussed the importance of our results, by comparing our approach
with previous work on random neural network models (SubSec\textcolor{blue}{.~}(\ref{subsec:Progress-with-Respect-to-Previous-Work-on-Bifurcation-Analysis})).
We also discussed the limitations of our technique (SubSec\textcolor{blue}{.~}(\ref{subsec:Limitations-of-Our-Approach})),
and the possibility to extend our work to the study of more complicated
kinds of neural dynamics (SubSec\textcolor{blue}{.~}(\ref{subsec:Future-Directions})).
To conclude, in the supplemental Python script ``Statistics.py'',
we implemented the comparison between our semi-analytical and numerical
results for arbitrary-size networks. Then, in the script ``Permanent.py'',
we implemented the comparison between the formula of the permanent
of statistically-homogeneous networks, that we introduced in SubSec.\textcolor{blue}{~}(\ref{subsec:The-Special-Case-of-Multi-Population-Networks-Composed-of-Statistically-Homogeneous-Populations}),
and the Balasubramanian-Bax-Franklin-Glynn algorithm. 

\section{Materials and Methods \label{sec:Materials-and-Methods}}

\subsection{The Network Model \label{subsec:The-Network-Model}}

We studied a recurrent neural network model composed of $N$ binary
neurons, whose state evolves in discrete time steps. The firing rate
of the $i$th neuron at time $t$ is represented by the binary variable
$\nu_{i}\left(t\right)\in\left\{ 0,1\right\} $, so that $\nu_{i}\left(t\right)=0$
if the neuron is not firing at time $t$, and $\nu_{i}\left(t\right)=1$
if it is firing at the maximum rate. We also defined $\boldsymbol{\nu}\left(t\right)\overset{\mathrm{def}}{=}\left(\begin{array}{cccc}
\nu_{0}\left(t\right), & \nu_{1}\left(t\right), & \cdots, & \nu_{N-1}\left(t\right)\end{array}\right)\in\left\{ 0,1\right\} ^{N}$, namely the vector containing the firing rates of all $N$ neurons
at time $t$.

If the neurons respond synchronously to the local fields $h_{i}\left(\boldsymbol{\nu}\left(t\right)\right)$,
the firing rates at the time instant $t+1$ are updated according
to the following activity-based equation (see e.g. \cite{Fasoli2018b}):

\begin{spacing}{0.8}
\begin{center}
{\small{}
\begin{equation}
\nu_{i}\left(t+1\right)=\mathscr{H}\left(h_{i}\left(\boldsymbol{\nu}\left(t\right)\right)-\theta_{i}\right),\quad h_{i}\left(\boldsymbol{\nu}\left(t\right)\right)\overset{\mathrm{def}}{=}\sum_{j=0}^{N-1}J_{i,j}\nu_{j}\left(t\right)+\mathfrak{I}_{i},\quad i=0,\cdots,N-1.\label{eq:synchronous-network-equations}
\end{equation}
}
\par\end{center}{\small \par}
\end{spacing}

\noindent As we anticipated, $N$ is the number of neurons in the
network, which in this work is supposed to be finite. Moreover, $\mathfrak{I}_{i}$
is a deterministic external input (i.e. an afferent stimulus) to the
$i$th neuron, while $\mathscr{H}\left(\cdot\right)$ is the Heaviside
step function:

\begin{spacing}{0.8}
\begin{center}
{\small{}
\begin{equation}
\mathscr{H}\left(h-\theta\right)=\begin{cases}
0, & \mathrm{if}\;\;h<\theta\\
\\
1, & \mathrm{if}\;\;h\geq\theta,
\end{cases}\label{eq:Heaviside-step-function}
\end{equation}
}
\par\end{center}{\small \par}
\end{spacing}

\noindent with deterministic firing threshold $\theta$.

$J_{i,j}$ is the (generally asymmetric) entry of the synaptic connectivity
matrix $J$, and represents the weight of the random and time-independent
synaptic connection from the $j$th (presynaptic) neuron to the $i$th
(postsynaptic) neuron. The randomness of the synaptic weights is \textit{quenched},
therefore a ``frozen'' connectivity matrix $J$ is randomly generated
at every realization of the network, according to the following formula:

\begin{spacing}{0.8}
\begin{center}
{\small{}
\begin{equation}
J_{i,j}=\begin{cases}
0, & \mathrm{if}\;\;T_{i,j}=0\\
\\
W_{i,j}, & \mathrm{if}\;\;T_{i,j}=1.
\end{cases}\label{eq:synaptic-weights}
\end{equation}
}
\par\end{center}{\small \par}
\end{spacing}

\noindent In Eq.~(\ref{eq:synaptic-weights}), $T_{i,j}$ is the
$\left(i,j\right)$th entry of the topology matrix $T$, so that $T_{i,j}=0$
if there is no synaptic connection from the $j$th neuron to the $i$th
neuron, and $T_{i,j}=1$ if the connection is present. The topology
of the network is generally random and asymmetric, and it depends
on the (arbitrary) entries $\mathcal{P}_{i,j}\in\left[0,1\right]$
of the connection probability matrix $\mathcal{P}$. In particular,
we supposed that $T_{i,j}=1$ with probability $\mathcal{P}_{i,j}$,
while $T_{i,j}=0$ with probability $1-\mathcal{P}_{i,j}$. Moreover,
in Eq.~(\ref{eq:synaptic-weights}) also the terms $W_{i,j}$ are
(generally asymmetric and non-identically distributed) random variables,
distributed according to marginal probability distributions $p_{W_{i,j}}$
(for simplicity, here we focused on continuous distributions, however
our calculations can be extended to discrete random variables, if
desired). In order to ensure the mathematical tractability of the
model, we supposed that the terms $W_{i,j}$ are statistically independent
from each other and from the variables $T_{i,j}$, and that the variables
$T_{i,j}$ are independent too.

On the other hand, if the neurons respond asynchronously to the local
fields, at each time instant only a single, randomly drawn neuron
$k$ is to undergo an update (see \cite{Fasoli2018b}):

\begin{spacing}{0.8}
\begin{center}
{\small{}
\begin{equation}
\begin{cases}
\nu_{i}\left(t+1\right)=\nu_{i}\left(t\right), & \forall i\neq k\\
\\
\nu_{i}\left(t+1\right)=\mathscr{H}\left(h_{i}\left(\boldsymbol{\nu}\left(t\right)\right)-\theta_{i}\right), & i=k,
\end{cases}\label{eq:asynchronous-network-equations}
\end{equation}
}
\par\end{center}{\small \par}
\end{spacing}

\noindent where the local field $h_{i}\left(\boldsymbol{\nu}\left(t\right)\right)$
is defined as in Eq.~(\ref{eq:synchronous-network-equations}). The
results that we derived in this paper are valid for both kinds of
network updates, since they generate identical bifurcation diagrams
of the stationary states for a given set of network parameters, as
we proved in \cite{Fasoli2018b}.

For simplicity, from now on we will represent the vector $\boldsymbol{\nu}\left(t\right)$
by the binary string $\nu_{0}\left(t\right)\nu_{1}\left(t\right)\cdots\nu_{N-1}\left(t\right)$,
obtained by concatenating the firing rates at time $t$. For example,
in a network composed of $N=6$ neurons, the vector $\boldsymbol{\nu}\left(t\right)=\left(\begin{array}{cccccc}
1, & 1, & 0, & 0, & 1, & 0\end{array}\right)$ will be represented by the string $110010$ (in this notation, no
multiplication is intended between the bits).

\subsection{Statistical Properties of the Network Model \label{subsec:Statistical-Properties-of-the-Network-Model}}

In this paper, we focused on the calculation of the statistical properties
of the stationary solutions of Eqs.~(\ref{eq:synchronous-network-equations})
and (\ref{eq:asynchronous-network-equations}), provided the probability
distribution of the entries $J_{i,j}$ of the connectivity matrix
is known. Therefore we did not consider the problem of storing/retrieving
some desired sequences of neural activity patterns.

Our formalism is based on a branch of statistics known as \textit{extreme
value theory} \cite{DeHaan2006}, which deals with the extreme deviations
from the median of probability distributions. Formally, extreme deviations
are described by the minimum and maximum of a set of random variables,
which correspond, respectively, to the smallest and largest \textit{order
statistics} of that set \cite{Vaughan1972,Bapat1989,Bapat1990,Hande1994}.

In this section, we derived semi-analytical formulas of the probability
distribution of the bifurcation points of the stationary states (SubSec.~(\ref{subsec:Probability-Distribution-of-the-Bifurcation-Points})),
of their mean multistability diagram (SubSec.~(\ref{subsec:Mean-Multistability-Diagram})),
of the probability that a state is stationary for a given combination
of stimuli (SubSec.~(\ref{subsec:Occurrence-Probability-of-the-Stationary-States-for-a-Given-Combination-of-Stimuli})),
and of the probability that a state is stationary regardless of the
stimuli (SubSec.~(\ref{subsec:Occurrence-Probability-of-the-Stationary-States-Regardless-of-the-Stimuli})).
We implemented these formulas in the ``Semi-Analytical Calculations''
section of the supplemental Python script ``Statistics.py''. Note
that our formulas are semi-analytical, in that they are expressed
in terms of 1D definite integrals containing the arbitrary probability
distributions $p_{W_{i,j}}$. In the Python script, these integrals
are calculated through numerical integration schemes. However, note
that generally the integrals may be calculated through analytical
approximations, while for some distributions exact formulas may also
exist, providing a fully-analytical description of the statistical
properties of the stationary states.

\subsubsection{Probability Distribution of the Bifurcation Points \label{subsec:Probability-Distribution-of-the-Bifurcation-Points}}

The \textit{multistability diagram} of the network provides a complete
picture of the relationship between the stationary solutions of the
network and a set of network parameters. In particular, in \cite{Fasoli2018b}
we studied how the degree of multistability (namely the number of
stationary solutions) of the firing-rate states and their symmetry
depend on the stimuli $\mathfrak{I}_{i}$, which represent our bifurcation
parameters. Given a network with $\mathfrak{P}$ distinct input currents
(i.e. $\mathfrak{I}_{i}\in\left\{ I_{0},\cdots,I_{\mathfrak{P}-1}\right\} \;\forall i$),
we defined $\Gamma_{I_{\alpha}}$ to be the set of neurons that share
the same external current $I_{\alpha}$ (namely $\Gamma_{I_{\alpha}}\overset{\mathrm{def}}{=}\left\{ i\in\left\{ 0,\cdots,N-1\right\} :\;\mathfrak{I}_{i}=I_{\alpha}\right\} $),
while $\Gamma_{I_{\alpha},u}^{\left(\boldsymbol{\nu}\right)}\overset{\mathrm{def}}{=}\left\{ i\in\Gamma_{I_{\alpha}}:\;\nu_{i}=u\right\} $,
for $u\in\left\{ 0,1\right\} $. Then in \cite{Fasoli2018b} we proved
that a given firing-rate state $\boldsymbol{\nu}$ is a stationary
solution of the network equations (\ref{eq:synchronous-network-equations})
and (\ref{eq:asynchronous-network-equations}) for every combination
of stimuli $\boldsymbol{I}\overset{\mathrm{def}}{=}\left(I_{0},\cdots,I_{\mathfrak{P}-1}\right)\in\mathscr{V}^{\left(\boldsymbol{\nu}\right)}=\mathcal{V}_{0}^{\left(\boldsymbol{\nu}\right)}\times\cdots\times\mathcal{V}_{\mathfrak{P}-1}^{\left(\boldsymbol{\nu}\right)}$,
where:

\begin{spacing}{0.8}
\begin{center}
{\small{}
\begin{align}
 & \mathcal{V}_{\alpha}^{\left(\boldsymbol{\nu}\right)}\overset{\mathrm{def}}{=}\begin{cases}
\left(-\infty,\Xi_{\alpha}^{\left(\boldsymbol{\nu}\right)}\right), & \mathrm{if}\;\;\Gamma_{I_{\alpha},1}^{\left(\boldsymbol{\nu}\right)}=\emptyset\\
\\
\left[\Lambda_{\alpha}^{\left(\boldsymbol{\nu}\right)},+\infty\right), & \mathrm{if}\;\;\Gamma_{I_{\alpha},0}^{\left(\boldsymbol{\nu}\right)}=\emptyset\\
\\
\left[\Lambda_{\alpha}^{\left(\boldsymbol{\nu}\right)},\Xi_{\alpha}^{\left(\boldsymbol{\nu}\right)}\right), & \mathrm{otherwise}
\end{cases}\nonumber \\
\nonumber \\
 & \Lambda_{\alpha}^{\left(\boldsymbol{\nu}\right)}\overset{\mathrm{def}}{=}\underset{i\in\Gamma_{I_{\alpha},1}^{\left(\boldsymbol{\nu}\right)}}{\max}\mathcal{I}_{i}^{\left(\boldsymbol{\nu}\right)},\quad\Xi_{\alpha}^{\left(\boldsymbol{\nu}\right)}\overset{\mathrm{def}}{=}\underset{i\in\Gamma_{I_{\alpha},0}^{\left(\boldsymbol{\nu}\right)}}{\min}\mathcal{I}_{i}^{\left(\boldsymbol{\nu}\right)}\label{eq:stationary-states-stimulus-range}\\
\nonumber \\
 & \mathcal{I}_{i}^{\left(\boldsymbol{\nu}\right)}\overset{\mathrm{def}}{=}\theta_{i}-\sum_{j=0}^{N-1}J_{i,j}\nu_{j}.\nonumber 
\end{align}
}
\par\end{center}{\small \par}
\end{spacing}

\noindent By calculating the hyperrectangles $\mathscr{V}^{\left(\boldsymbol{\nu}\right)}$
for every $\boldsymbol{\nu}$, we obtain a complete picture of the
relationship between the stationary states and the set of stimuli.
If the hyperrectangles corresponding to $\mathcal{M}$ different states
$\boldsymbol{\nu}$ overlap, the overlapping region has multistability
degree $\mathcal{M}$ (i.e. for combinations of stimuli lying in that
region, the network has $\mathcal{M}$ distinct stationary states).
A stationary state loses its stability at the boundaries $\Lambda_{\alpha}^{\left(\boldsymbol{\nu}\right)}$
and $\Xi_{\alpha}^{\left(\boldsymbol{\nu}\right)}$, turning into
another stationary state or an oscillation. Therefore $\Lambda_{\alpha}^{\left(\boldsymbol{\nu}\right)}$
and $\Xi_{\alpha}^{\left(\boldsymbol{\nu}\right)}$ represent the
\textit{bifurcation points} of the stationary solution $\boldsymbol{\nu}$
of Eqs.~(\ref{eq:synchronous-network-equations}) and (\ref{eq:asynchronous-network-equations}).

In what follows, we derived the probability density functions of the
bifurcation points (note that, for simplicity, from now on we will
omit the superscript $\left(\boldsymbol{\nu}\right)$ in all the formulas).
Since the random variables $\mathcal{I}_{i}$ are independent for
a given firing-rate state $\boldsymbol{\nu}$ (as a consequence of
the independence of the synaptic weights $J_{i,j}$), if we call $\mathrm{per}\left(\cdot\right)$
the matrix permanent and $F_{\mathcal{I}_{i}}\left(x\right)\overset{\mathrm{def}}{=}\int_{-\infty}^{x}p_{\mathcal{I}_{i}}\left(y\right)dy$
the cumulative distribution function of $\mathcal{I}_{i}$, then from
Eq.~(\ref{eq:stationary-states-stimulus-range}) we obtain that $\Lambda_{\alpha}^{\left(\boldsymbol{\nu}\right)}$
and $\Xi_{\alpha}^{\left(\boldsymbol{\nu}\right)}$ are distributed
according to the following order statistics \cite{Vaughan1972,Bapat1989,Bapat1990,Hande1994}:

\begin{spacing}{0.8}
\begin{center}
{\small{}
\begin{align}
p_{\Lambda_{\alpha}}\left(x\right)= & \frac{1}{\left(\gamma_{\alpha,1}-1\right)!}\mathrm{per}\left(\left[\begin{array}{cc}
\boldsymbol{p}_{\alpha,1}\left(x\right), & \boldsymbol{F}_{\alpha,1}^{\left(\gamma_{\alpha,1}-1\right)}\left(x\right)\end{array}\right]\right)\nonumber \\
\label{eq:probability-distribution-bifurcation-points}\\
p_{\Xi_{\alpha}}\left(x\right)= & \frac{1}{\left(\gamma_{\alpha,0}-1\right)!}\mathrm{per}\left(\left[\begin{array}{cc}
\boldsymbol{p}_{\alpha,0}\left(x\right), & \mathbb{I}_{\gamma_{\alpha,0},\gamma_{\alpha,0}-1}-\boldsymbol{F}_{\alpha,0}^{\left(\gamma_{\alpha,0}-1\right)}\left(x\right)\end{array}\right]\right),\nonumber 
\end{align}
}
\par\end{center}{\small \par}
\end{spacing}

\noindent where $\gamma_{\alpha,u}\overset{\mathrm{def}}{=}\left|\Gamma_{I_{\alpha},u}\right|$,
while $\left[\begin{array}{cc}
\boldsymbol{p}_{\alpha,1}\left(x\right), & \boldsymbol{F}_{\alpha,1}^{\left(\gamma_{\alpha,1}-1\right)}\left(x\right)\end{array}\right]$ and $\left[\begin{array}{cc}
\boldsymbol{p}_{\alpha,0}\left(x\right), & \mathbb{I}_{\gamma_{\alpha,0},\gamma_{\alpha,0}-1}-\boldsymbol{F}_{\alpha,0}^{\left(\gamma_{\alpha,0}-1\right)}\left(x\right)\end{array}\right]$ are $\gamma_{\alpha,1}\times\gamma_{\alpha,1}$ and $\gamma_{\alpha,0}\times\gamma_{\alpha,0}$
matrices respectively, $\boldsymbol{p}_{\alpha,u}\left(x\right)\overset{\mathrm{def}}{=}\left[p_{\mathcal{I}_{i}}\left(x\right)\right]_{i\in\Gamma_{I_{\alpha},u}}$
and $\boldsymbol{F}_{\alpha,u}\left(x\right)\overset{\mathrm{def}}{=}\left[F_{\mathcal{I}_{i}}\left(x\right)\right]_{i\in\Gamma_{I_{\alpha},u}}$
are $\gamma_{\alpha,u}\times1$ column vectors, $\boldsymbol{F}_{\alpha,u}^{\left(v\right)}\left(x\right)\overset{\mathrm{def}}{=}\underset{v-\mathrm{times}}{\left[\underbrace{\begin{array}{ccc}
\boldsymbol{F}_{\alpha,u}\left(x\right), & \cdots, & \boldsymbol{F}_{\alpha,u}\left(x\right)\end{array}}\right]}$ is a $\gamma_{\alpha,u}\times v$ matrix, and $\mathbb{I}_{\gamma_{\alpha,0},\gamma_{\alpha,0}-1}$
is the $\gamma_{\alpha,0}\times\left(\gamma_{\alpha,0}-1\right)$
all-ones matrix. Note that in the supplemental Python script ``Statistics.py''
the permanent is calculated by means of the Balasubramanian-Bax-Franklin-Glynn
(BBFG) formula, which is the fastest known algorithm for the numerical
calculation of the permanent of arbitrary matrices \cite{Balasubramanian1980,Bax1996,Bax1998,Glynn2010}.

According to Eq.~(\ref{eq:probability-distribution-bifurcation-points}),
in order to complete the derivation of the probability densities $p_{\Lambda_{\alpha}}$
and $p_{\Xi_{\alpha}}$, we need to evaluate the probability density
and the cumulative distribution function of $\mathcal{I}_{i}$. By
defining $S_{i}\overset{\mathrm{def}}{=}\sum_{j=0}^{N-1}J_{i,j}\nu_{j}=\sum_{j\in R}J_{i,j}$
and $R\overset{\mathrm{def}}{=}\left\{ j\in\left\{ 0,\cdots,N-1\right\} :\;\nu_{j}=1\right\} $,
from the definition of $\mathcal{I}_{i}$ in Eq.~(\ref{eq:stationary-states-stimulus-range})
it follows that:

\begin{spacing}{0.8}
\begin{center}
\textit{\small{}
\begin{equation}
p_{\mathcal{I}_{i}}\left(x\right)=p_{S_{i}}\left(\theta_{i}-x\right).\label{eq:probability-distribution-of-I}
\end{equation}
}
\par\end{center}{\small \par}
\end{spacing}

\noindent Since the synaptic weights $J_{i,j}$ are independent by
hypothesis, the probability distribution of $S_{i}$ can be calculated
through the convolution formula:

\begin{spacing}{0.8}
\begin{center}
\textit{\small{}
\begin{equation}
p_{S_{i}}\left(\theta_{i}-x\right)=\left(\name_{j\in R}p_{J_{i,j}}\right)\left(\theta_{i}-x\right),\label{eq:convolution-0}
\end{equation}
}
\par\end{center}{\small \par}
\end{spacing}

\noindent According to Eq.~(\ref{eq:synaptic-weights}):

\begin{spacing}{0.8}
\begin{center}
\textit{\small{}
\begin{equation}
p_{J_{i,j}}\left(x\right)=\mathcal{P}_{i,j}p_{W_{i,j}}\left(x\right)+\left(1-\mathcal{P}_{i,j}\right)\delta\left(x\right),\label{eq:synaptic-weights-distribution}
\end{equation}
}
\par\end{center}{\small \par}
\end{spacing}

\noindent where $\delta\left(\cdot\right)$ is the Dirac delta function.
Then, from Eqs.~(\ref{eq:convolution-0}) and (\ref{eq:synaptic-weights-distribution}),
it follows that:

\begin{spacing}{0.8}
\begin{center}
\textit{\small{}
\begin{align}
p_{S_{i}}\left(\theta_{i}-x\right)= & a_{i}\left(\theta_{i}-x\right)+b_{i}\delta\left(\theta_{i}-x\right)\nonumber \\
\nonumber \\
a_{i}\left(x\right)= & \sum_{\mathscr{S}\in\mathbb{P}\left(R\right)\backslash\emptyset}\left[\prod_{j\in\mathscr{S}}\mathcal{P}_{i,j}\right]\left[\prod_{j\in R\backslash\mathscr{S}}\left(1-\mathcal{P}_{i,j}\right)\right]\left[\left(\name_{j\in\mathscr{S}}p_{W_{i,j}}\right)\left(x\right)\right]\nonumber \\
\nonumber \\
b_{i}= & \prod_{j\in R}\left(1-\mathcal{P}_{i,j}\right),\label{eq:convolution-1}
\end{align}
}
\par\end{center}{\small \par}
\end{spacing}

\noindent where $\mathbb{P}\left(R\right)$ represents the power set
of $R$. Finally, we obtain $p_{\mathcal{I}_{i}}\left(x\right)$ from
Eqs.~(\ref{eq:probability-distribution-of-I}) and (\ref{eq:convolution-1}),
while the corresponding cumulative distribution function is:

\begin{spacing}{0.8}
\begin{center}
\textit{\small{}
\begin{equation}
F_{\mathcal{I}_{i}}\left(x\right)=\int_{-\infty}^{x}a_{i}\left(\theta_{i}-y\right)dy+b_{i}\mathscr{H}\left(x-\theta_{i}\right).\label{eq:cumulative-distribution-function-of-I}
\end{equation}
}
\par\end{center}{\small \par}
\end{spacing}

\noindent Note that the definite integral in Eq.~(\ref{eq:cumulative-distribution-function-of-I})
depends on the probability distribution $p_{W_{i,j}}$, which is arbitrary.
For this reason, we did not provide any analytical expression of this
integral, though exact formulas exist for specific distributions $p_{W_{i,j}}$,
e.g. when $W_{i,j}$ are normally distributed (in the supplemental
Python script ``Statistics.py'', the distribution $p_{W_{i,j}}$
is defined by the user, and the integrals are calculated by means
of numerical integration schemes).

Now we have all the ingredients for calculating the probability distributions
of the bifurcation points from Eq.~(\ref{eq:probability-distribution-bifurcation-points}).
Note, however, that this formula cannot be used in its current form,
because it involves the ill-defined product between the Dirac delta
function (which is contained in the probability density $p_{\mathcal{I}_{i}}$)
and a discontinuous function (i.e. $F_{\mathcal{I}_{i}}$, which jumps
at $x=\theta_{i}$). To see that this product is generally ill-defined,
consider for example the probability density of $\underset{j}{\mathrm{min}}\left(X_{j}\right)$,
in the case when $X_{j}$ are $n$ independent and identically distributed
random variables:
\begin{spacing}{0.8}
\begin{center}
{\small{}
\begin{equation}
p_{\underset{j}{\mathrm{min}}\left(X_{j}\right)}\left(x\right)=\frac{1}{\left(n-1\right)!}\mathrm{per}\left(\left[\begin{array}{c}
p_{X}\left(x\right)\\
\vdots\\
p_{X}\left(x\right)
\end{array}\underset{\left(n-1\right)-\mathrm{times}}{\underbrace{\begin{array}{ccc}
F_{X}\left(x\right) & \cdots & F_{X}\left(x\right)\\
\vdots & \ddots & \vdots\\
F_{X}\left(x\right) & \cdots & F_{X}\left(x\right)
\end{array}}}\right]\right)=np_{X}\left(x\right)F_{X}^{n-1}\left(x\right).\label{eq:probability-density-iid-variables}
\end{equation}
}
\par\end{center}{\small \par}
\end{spacing}

\noindent If $F_{X}\left(x\right)$ is absolutely continuous, we can
prove that $p_{\underset{j}{\mathrm{min}}\left(X_{j}\right)}\left(x\right)$,
as given by Eq.~(\ref{eq:probability-density-iid-variables}), is
correctly normalized:
\begin{spacing}{0.8}
\begin{center}
{\small{}
\[
\int_{-\infty}^{+\infty}np_{X}\left(x\right)F_{X}^{n-1}\left(x\right)dx=\int_{-\infty}^{+\infty}n\frac{dF_{X}\left(x\right)}{dx}F_{X}^{n-1}\left(x\right)dx=\int_{0}^{1}nF_{X}^{n-1}dF_{X}=\left.n\frac{1}{n}F_{X}^{n}\right|_{0}^{1}=1.
\]
}
\par\end{center}{\small \par}
\end{spacing}

\noindent However, in the limit $p_{X}\left(x\right)\rightarrow\delta\left(x\right)$,
the cumulative distribution function $F_{X}\left(x\right)$ is discontinuous
at $x=0$, and we get:
\begin{spacing}{0.8}
\begin{center}
{\small{}
\[
\int_{-\infty}^{+\infty}np_{X}\left(x\right)F_{X}^{n-1}\left(x\right)dx\rightarrow\int_{-\infty}^{+\infty}n\delta\left(x\right)\mathscr{H}^{n-1}\left(x\right)dx.
\]
}
\par\end{center}{\small \par}
\end{spacing}

\noindent If now we attempt to apply the famous formula:
\begin{spacing}{0.8}
\begin{center}
{\small{}
\[
\int_{-\infty}^{+\infty}\delta\left(x\right)f\left(x\right)dx=f\left(0\right),
\]
}
\par\end{center}{\small \par}
\end{spacing}

\noindent we get that the integral equals $n\mathscr{H}^{n-1}\left(0\right)=n>1$,
therefore $p_{\underset{j}{\mathrm{min}}\left(X_{j}\right)}\left(x\right)$
is not properly normalized. The same problem occurs for $\underset{j}{\mathrm{max}}\left(X_{j}\right)$.

To fix it, consider the general case when $X$ is a mixture of continuous
and discrete random variables. Its probability density can be decomposed
as follows:

\begin{spacing}{0.8}
\begin{center}
{\small{}
\begin{equation}
p_{X}\left(x\right)=p_{X^{c}}\left(x\right)+\sum_{q\in D}\left[F_{X}\left(x_{q}\right)-\underset{x\rightarrow x_{q}^{-}}{\lim}F_{X}\left(x\right)\right]\delta\left(x-x_{q}\right).\label{eq:probability-distribution-decomposition}
\end{equation}
}
\par\end{center}{\small \par}
\end{spacing}

\noindent In Eq.~(\ref{eq:probability-distribution-decomposition}),
$p_{X^{c}}$ is the component of $p_{X}$ that describes the statistical
behavior of the continuous values of $X$. Moreover, $\left\{ x_{q}\right\} _{q\in D}$
represents the set of the discrete values of $X$, at which the cumulative
distribution function $F_{X}$ is (possibly) discontinuous. In the
specific case when $X=\Lambda_{\alpha}$ or $X=\Xi_{\alpha}$, by
comparing Eq.~(\ref{eq:probability-distribution-decomposition})
with Eqs.~(\ref{eq:probability-distribution-bifurcation-points}),
(\ref{eq:probability-distribution-of-I}) and (\ref{eq:convolution-1}),
we get:

\begin{spacing}{0.8}
\begin{center}
{\small{}
\begin{align}
p_{\Lambda_{\alpha}^{c}}\left(x\right)= & \frac{1}{\left(\gamma_{\alpha,1}-1\right)!}\mathrm{per}\left(\left[\begin{array}{cc}
\boldsymbol{a}_{\alpha,1}\left(\boldsymbol{\theta}-\boldsymbol{x}\right), & \boldsymbol{F}_{\alpha,1}^{\left(\gamma_{\alpha,1}-1\right)}\left(x\right)\end{array}\right]\right)\nonumber \\
\label{eq:continuous-components}\\
p_{\Xi_{\alpha}^{c}}\left(x\right)= & \frac{1}{\left(\gamma_{\alpha,0}-1\right)!}\mathrm{per}\left(\left[\begin{array}{cc}
\boldsymbol{a}_{\alpha,0}\left(\boldsymbol{\theta}-\boldsymbol{x}\right), & \mathbb{I}_{\gamma_{\alpha,0},\gamma_{\alpha,0}-1}-\boldsymbol{F}_{\alpha,0}^{\left(\gamma_{\alpha,0}-1\right)}\left(x\right)\end{array}\right]\right),\nonumber 
\end{align}
}
\par\end{center}{\small \par}
\end{spacing}

\noindent where $\boldsymbol{a}_{\alpha,u}\left(\boldsymbol{\theta}-\boldsymbol{x}\right)\overset{\mathrm{def}}{=}\left[a_{i}\left(\theta_{i}-x\right)\right]_{i\in\Gamma_{I_{\alpha},u}}$
is a $\gamma_{\alpha,u}\times1$ column vector. Moreover, $D=\Gamma_{I_{\alpha},1}$
and $D=\Gamma_{I_{\alpha},0}$ for $\Lambda_{\alpha}$ and $\Xi_{\alpha}$
respectively, while $x_{q}=\theta_{q}$. According to Eq.~(\ref{eq:probability-distribution-decomposition}),
we need also to evaluate the cumulative distribution functions of
$\Lambda_{\alpha}$ and $\Xi_{\alpha}$. By following \cite{Bapat1990},
we get:

\begin{spacing}{0.8}
\begin{center}
{\small{}
\begin{align}
F_{\Lambda_{\alpha}}\left(x\right)= & \frac{1}{\gamma_{\alpha,1}!}\mathrm{per}\left(\left[\boldsymbol{F}_{\alpha,1}^{\left(\gamma_{\alpha,1}\right)}\left(x\right)\right]\right)\nonumber \\
\label{eq:cumulative-distribution-functions}\\
F_{\Xi_{\alpha}}\left(x\right)= & \sum_{n=1}^{\gamma_{\alpha,0}}\frac{1}{n!\left(\gamma_{\alpha,0}-n\right)!}\mathrm{per}\left(\left[\begin{array}{cc}
\boldsymbol{F}_{\alpha,0}^{\left(n\right)}\left(x\right), & \mathbb{I}_{\gamma_{\alpha,0},\gamma_{\alpha,0}-n}-\boldsymbol{F}_{\alpha,0}^{\left(\gamma_{\alpha,0}-n\right)}\left(x\right)\end{array}\right]\right).\nonumber 
\end{align}
}
\par\end{center}{\small \par}
\end{spacing}

\noindent We observe that Eqs.~(\ref{eq:probability-distribution-decomposition}),
(\ref{eq:continuous-components}) and (\ref{eq:cumulative-distribution-functions})
do not depend anymore on the ill-defined product between the Dirac
delta distribution and the Heaviside step function. These formulas
will be used in the next subsections to calculate the mean multistability
diagram of the network (SubSec.~(\ref{subsec:Mean-Multistability-Diagram})),
the probability that a firing-rate state is stationary for a given
combination of stimuli (SubSec.~(\ref{subsec:Occurrence-Probability-of-the-Stationary-States-for-a-Given-Combination-of-Stimuli})),
and the probability that a state is stationary regardless of the stimuli
(SubSec.~(\ref{subsec:Occurrence-Probability-of-the-Stationary-States-Regardless-of-the-Stimuli})).

\subsubsection{Mean Multistability Diagram \label{subsec:Mean-Multistability-Diagram}}

The mean multistability diagram is the plot of the bifurcation points
$\Lambda_{\alpha}$ and $\Xi_{\alpha}$, averaged over the network
realizations. The mean bifurcation points $\left\langle \Lambda_{\alpha}\right\rangle $
and $\left\langle \Xi_{\alpha}\right\rangle $ (where the brackets
$\left\langle \cdot\right\rangle $ represent the statistical mean
over the network realizations) correspond to the values of the stimulus
$I_{\alpha}$ at which a given firing-rate state $\boldsymbol{\nu}$
loses its stability on average, turning into a different stationary
state or an oscillatory solution. We propose two different approaches
for evaluating the mean bifurcation points, which we implemented in
the supplemental Python script ``Statistics.py''.

The first method is based on Eq.~(\ref{eq:probability-distribution-decomposition}),
from which we obtain:

\begin{spacing}{0.8}
\begin{center}
{\small{}
\begin{equation}
\left\langle X\right\rangle =\int_{-\infty}^{+\infty}xp_{X}\left(x\right)dx=\int_{-\infty}^{+\infty}xp_{X^{c}}\left(x\right)dx+\sum_{q\in D}x_{q}\left[F_{X}\left(x_{q}\right)-\underset{x\rightarrow x_{q}^{-}}{\lim}F_{X}\left(x\right)\right],\label{eq:mean_method_1}
\end{equation}
}
\par\end{center}{\small \par}
\end{spacing}

\noindent for $X=\Lambda_{\alpha}$ and $X=\Xi_{\alpha}$. The cumulative
distribution function $F_{X}$ in Eq.~(\ref{eq:mean_method_1}) is
calculated by means of Eq.~(\ref{eq:cumulative-distribution-functions}),
while the function $p_{X^{c}}$ is given by Eq.~(\ref{eq:continuous-components}).

The second method takes advantage of the following formula:

\begin{spacing}{0.8}
\begin{center}
{\small{}
\[
\left\langle X^{z}\right\rangle =\int_{-\infty}^{+\infty}x^{z}dF_{X}\left(x\right)=z\int_{0}^{+\infty}x^{z-1}\left[1-F_{X}\left(x\right)+\left(-1\right)^{z}F_{X}\left(-x\right)\right]dx,
\]
}
\par\end{center}{\small \par}
\end{spacing}

\noindent where the second equality is obtained by integrating the
Lebesgue-Stieltjes integral by parts. After some algebra, in the special
case $z=1$ we get:

\begin{spacing}{0.8}
\begin{center}
{\small{}
\begin{equation}
\left\langle X\right\rangle =\int_{-\infty}^{+\infty}\left[\mathscr{H}\left(x\right)-F_{X}\left(x\right)\right]dx,\label{eq:mean_method_2}
\end{equation}
}
\par\end{center}{\small \par}
\end{spacing}

\noindent where $F_{X}$ is given again by Eq.~(\ref{eq:cumulative-distribution-functions}).
By running both methods in the supplemental Python script, the reader
can easily check that Eqs.~(\ref{eq:mean_method_1}) and (\ref{eq:mean_method_2})
provide identical results for $\left\langle \Lambda_{\alpha}\right\rangle $
and $\left\langle \Xi_{\alpha}\right\rangle $, apart from rounding
errors.

It is important to observe that the multistability diagram shows only
those stability regions for which $\left\langle \Lambda_{\alpha}\right\rangle <\left\langle \Xi_{\alpha}\right\rangle $
for every $\alpha$, because if this condition is not satisfied, the
state $\boldsymbol{\nu}$ is not stationary on average for any combination
of stimuli. Moreover, beyond multistability, the diagram provides
also a complete picture of spontaneous symmetry-breaking of the stationary
solutions of the firing rates. Spontaneous symmetry-breaking occurs
whenever neurons in homogeneous populations fire at different rates,
despite the symmetry of the underlying neural equations. We define
the population function $\mathfrak{p}\left(\cdot\right)$ that maps
the neuron index $i\in\left\{ 0,\cdots,N-1\right\} $ to the index
$\alpha$ of the population the neuron $i$ belongs to, so that $\mathfrak{p}\left(i\right)=\alpha$.
Then, in a single network realization, a population $\alpha$ is said
to be \textit{homogeneous} if the sum $\mathscr{S}_{i}^{\left(\beta\right)}\overset{\mathrm{def}}{=}\sum_{k:\,\mathfrak{p}\left(k\right)=\beta}J_{i,k}$,
the firing threshold $\theta_{i}$ and the external stimulus $\mathfrak{I}_{i}$
do not depend on the index $i$, for every index $i$ such that $\mathfrak{p}\left(i\right)=\alpha$
(see \cite{Fasoli2018b}). However, in the present article we studied
network statistics across realizations. For this reason, the homogeneity
of a neural population should be defined in a statistical sense, namely
by checking whether the probability distribution of $\mathscr{S}_{i}^{\left(\beta\right)}$
does not depend on the index $i$, for every neuron $i$ in the population
$\alpha$. Whenever the neurons in a population show heterogeneous
firing rates despite the homogeneity condition is satisfied, we say
that the symmetry of that population is spontaneously broken. In order
to check whether the probability distribution of $\mathscr{S}_{i}^{\left(\beta\right)}$
is population-dependent, it is possible to calculate numerically the
Kullback-Leibler divergence $\mathcal{D}_{\mathrm{KL}}\left(\mathscr{S}_{i}^{\left(\beta\right)}\parallel\mathscr{S}_{j}^{\left(\beta\right)}\right)$
between all the pairs of neurons $i,\;j$ that belong to the same
population $\alpha$. However, in the supplemental script ``Statistics.py'',
we checked the statistical homogeneity of the neural populations in
a simpler and computationally more efficient way, though our approach
is less general than that based on the Kullback-Leibler divergence.
Our method relies on the assumption that a small number of moments
of $\mathscr{S}_{i}^{\left(\beta\right)},\;\mathscr{S}_{j}^{\left(\beta\right)}$,
for example just the mean and the variance:
\begin{spacing}{0.8}
\begin{center}
{\small{}
\begin{align*}
\left\langle \mathscr{S}_{i}^{\left(\beta\right)}\right\rangle = & \sum_{k:\,\mathfrak{p}\left(k\right)=\beta}\mathcal{P}_{i,k}\left\langle W_{i,k}\right\rangle \\
\\
\mathrm{Var}\left(\mathscr{S}_{i}^{\left(\beta\right)}\right)= & \sum_{k:\,\mathfrak{p}\left(k\right)=\beta}\left(\mathcal{P}_{i,k}\left\langle W_{i,k}^{2}\right\rangle -\mathcal{P}_{i,j}^{2}\left\langle W_{i,k}\right\rangle ^{2}\right),
\end{align*}
}
\par\end{center}{\small \par}
\end{spacing}

\noindent are sufficient for discriminating between the probability
distributions of the two random variables. In other words, we assumed
that if $\left\langle \mathscr{S}_{i}^{\left(\beta\right)}\right\rangle \neq\left\langle \mathscr{S}_{j}^{\left(\beta\right)}\right\rangle $
and/or $\mathrm{Var}\left(\mathscr{S}_{i}^{\left(\beta\right)}\right)\neq\mathrm{Var}\left(\mathscr{S}_{j}^{\left(\beta\right)}\right)$,
then $\mathscr{S}_{i}^{\left(\beta\right)},\;\mathscr{S}_{j}^{\left(\beta\right)}$
are differently distributed, and therefore the neural population $\alpha$
is statistically heterogeneous~\footnote{\noindent Note, however, that the probability distribution of a scalar
random variable with finite moments at all orders, generally is not
uniquely determined by the sequence of moments. It follows that there
exist (rare) cases of differently distributed random variables that
share the same sequence of moments. For this reason, the moments are
not always sufficient for discriminating between two probability distributions.
Note also that a sufficient condition for the sequence of moments
to uniquely determine the random variable is that the moment generating
function has positive radius of convergence (see Thm.~(30.1) in \cite{Billingsley1995}).}.

\subsubsection{Occurrence Probability of the Stationary States for a Given Combination
of Stimuli \label{subsec:Occurrence-Probability-of-the-Stationary-States-for-a-Given-Combination-of-Stimuli}}

In this subsection we calculated the probability that a given firing-rate
state $\boldsymbol{\nu}$ is stationary, for a fixed combination of
stimuli. According to Eq.~(\ref{eq:stationary-states-stimulus-range}),
$\boldsymbol{\nu}$ is stationary for every $\boldsymbol{I}\in\mathscr{V}$.
Since the boundaries of $\mathscr{V}$ (namely the functions $\Lambda_{\alpha}$
and $\Xi_{\alpha}$) are random variables, it follows that the probability
that the firing-rate state $\boldsymbol{\nu}$ is stationary, for
a fixed combination of stimuli $\widehat{\boldsymbol{I}}\overset{\mathrm{def}}{=}\left(\widehat{I}_{0},\cdots,\widehat{I}_{\mathfrak{P}-1}\right)$,
is $P\left(\widehat{\boldsymbol{I}}\in\mathscr{V}=\mathcal{V}_{0}\times\cdots\times\mathcal{V}_{\mathfrak{P}-1}\right)$.
Since $\Lambda_{\alpha}$, for a given firing rate $\boldsymbol{\nu}$
and for $\alpha\in\left\{ 0,\cdots,\mathfrak{P}-1\right\} $, are
independent variables (and the same for the variables $\Xi_{\alpha}$),
it follows that $P\left(\widehat{\boldsymbol{I}}\in\mathscr{V}\right)$
can be factored out into the product of the probabilities $P\left(\widehat{I}_{\alpha}\in\mathcal{V}_{\alpha}\right)$.
In particular, whenever $\Gamma_{I_{\alpha},1}=\emptyset$, from Eq.~(\ref{eq:stationary-states-stimulus-range})
we see that $\boldsymbol{\nu}$ is stationary for every $I_{\alpha}<\Xi_{\alpha}$.
It follows that, in this case, $P\left(\widehat{I}_{\alpha}\in\mathcal{V}_{\alpha}\right)=P\left(\widehat{I}_{\alpha}<\Xi_{\alpha}\right)=\int_{\widehat{I}_{\alpha}}^{+\infty}p_{\Xi_{\alpha}}\left(x\right)dx=1-F_{\Xi_{\alpha}}\left(\widehat{I}_{\alpha}\right)$.
On the other hand, whenever $\Gamma_{I_{\alpha},0}=\emptyset$, the
state $\boldsymbol{\nu}$ is stationary for every $I_{\alpha}\geq\Lambda_{\alpha}$,
so that $P\left(\widehat{I}_{\alpha}\in\mathcal{V}_{\alpha}\right)=P\left(\widehat{I}_{\alpha}\geq\Lambda_{\alpha}\right)=\int_{-\infty}^{\widehat{I}_{\alpha}}p_{\Lambda_{\alpha}}\left(x\right)dx=F_{\Lambda_{\alpha}}\left(\widehat{I}_{\alpha}\right)$.
In all the other cases, $\boldsymbol{\nu}$ is stationary for every
$\Lambda_{\alpha}\leq I_{\alpha}<\Xi_{\alpha}$. This condition can
be decomposed as $\left(I_{\alpha}\geq\Lambda_{\alpha}\right)\wedge\left(I_{\alpha}<\Xi_{\alpha}\right)$,
and since $\Gamma_{I_{\alpha},0}\cap\Gamma_{I_{\alpha},1}=\emptyset$,
the random variables $\Lambda_{\alpha}$ e $\Xi_{\alpha}$ are independent,
so that $P\left(\widehat{I}_{\alpha}\in\mathcal{V}_{\alpha}\right)=P\left(\widehat{I}_{\alpha}\geq\Lambda_{\alpha}\right)P\left(\widehat{I}_{\alpha}<\Xi_{\alpha}\right)=F_{\Lambda_{\alpha}}\left(\widehat{I}_{\alpha}\right)\left[1-F_{\Xi_{\alpha}}\left(\widehat{I}_{\alpha}\right)\right]$.
Therefore, to summarize, the probability that the firing-rate state
$\boldsymbol{\nu}$ is stationary, for a fixed combination of stimuli
$\widehat{\boldsymbol{I}}$, is:
\begin{spacing}{0.8}
\begin{center}
{\small{}
\begin{align}
 & P\left(\widehat{\boldsymbol{I}}\in\mathscr{V}\right)=\prod_{\alpha=0}^{\mathfrak{P}-1}P\left(\widehat{I}_{\alpha}\in\mathcal{V}_{\alpha}\right)\label{eq:probability-for-fixed-stimuli}\\
\nonumber \\
 & P\left(\widehat{I}_{\alpha}\in\mathcal{V}_{\alpha}\right)=\begin{cases}
1-F_{\Xi_{\alpha}}\left(\widehat{I}_{\alpha}\right), & \mathrm{if}\;\;\Gamma_{I_{\alpha},1}=\emptyset\\
\\
F_{\Lambda_{\alpha}}\left(\widehat{I}_{\alpha}\right), & \mathrm{if}\;\;\Gamma_{I_{\alpha},0}=\emptyset\\
\\
F_{\Lambda_{\alpha}}\left(\widehat{I}_{\alpha}\right)\left[1-F_{\Xi_{\alpha}}\left(\widehat{I}_{\alpha}\right)\right], & \mathrm{otherwise}.
\end{cases}\nonumber 
\end{align}
}
\par\end{center}{\small \par}
\end{spacing}

\noindent Note that $P\left(\widehat{\boldsymbol{I}}\in\mathscr{V}\right)$
can be equivalently interpreted as the conditional probability $P\left(\boldsymbol{\nu}|\boldsymbol{\nu},\widehat{\boldsymbol{I}}\right)$,
and that generally $\sum_{\boldsymbol{\nu}\in\left\{ 0,1\right\} ^{N}}P\left(\boldsymbol{\nu}|\boldsymbol{\nu},\widehat{\boldsymbol{I}}\right)\neq1$,
therefore $P\left(\widehat{\boldsymbol{I}}\in\mathscr{V}\right)$
is not normalized over the set of the possible $2^{N}$ firing-rate
states.

\subsubsection{Occurrence Probability of the Stationary States Regardless of the
Stimuli \label{subsec:Occurrence-Probability-of-the-Stationary-States-Regardless-of-the-Stimuli}}

In this subsection we calculated the probability to observe a given
firing-rate state $\boldsymbol{\nu}$ in the whole multistability
diagram of a single network realization, namely the probability that
the state $\boldsymbol{\nu}$ is stationary regardless of the specific
combination of stimuli to the network. In other words, this represents
the probability that $\boldsymbol{\nu}$ is stationary for at least
one combination of stimuli. The firing-rate state $\boldsymbol{\nu}$
is observed in the multistability diagram only if its corresponding
hyperrectangle $\mathscr{V}$ has positive hypervolume $\mathrm{vol}\left(\mathscr{V}\right)$.
Since $\mathscr{V}=\mathcal{V}_{0}\times\cdots\times\mathcal{V}_{\mathfrak{P}-1}$,
it follows that $\mathrm{vol}\left(\mathscr{V}\right)=\prod_{\alpha=0}^{\mathfrak{P}-1}\mathrm{len}\left(\mathcal{V}_{\alpha}\right)>0$
only if $\mathrm{len}\left(\mathcal{V}_{\alpha}\right)>0\;\forall\alpha$,
where $\mathrm{len}\left(\mathcal{V}_{\alpha}\right)$ represents
the length of the interval $\mathcal{V}_{\alpha}$~\footnote{Note that in Sec.~(\ref{sec:Results}) we consider an example of
neural network model with $\mathfrak{P}=2$, therefore in that case
$\prod_{\alpha=0}^{\mathfrak{P}-1}\mathrm{len}\left(\mathcal{V}_{\alpha}\right)$
represents the area of the rectangles in the stimuli space. Nevertheless,
to avoid confusion, we will continue to use the general notation $\mathrm{vol}\left(\mathscr{V}\right)$.}. In particular, when $\Gamma_{I_{\alpha},0}=\emptyset$ (respectively
$\Gamma_{I_{\alpha},1}=\emptyset$), from Eq.~(\ref{eq:stationary-states-stimulus-range})
we get $\mathrm{len}\left(\mathcal{V}_{\alpha}\right)=\infty$ for
every $\Lambda_{\alpha}$ (respectively $\Xi_{\alpha}$), or in other
words $P\left(\mathrm{len}\left(\mathcal{V}_{\alpha}\right)>0\right)=1$.
On the other hand, when $\Gamma_{I_{\alpha},0},\;\Gamma_{I_{\alpha},1}\neq\emptyset$,
according to Eq.~(\ref{eq:stationary-states-stimulus-range}) we
obtain $\mathrm{len}\left(\mathcal{V}_{\alpha}\right)=\Xi_{\alpha}-\Lambda_{\alpha}$.
Since $\Lambda_{\alpha}$ and $\Xi_{\alpha}$ are independent for
a given $\boldsymbol{\nu}$, we can write:
\begin{spacing}{0.8}
\begin{center}
{\small{}
\[
p_{\mathrm{len}\left(\mathcal{V}_{\alpha}\right)}\left(x\right)=\int_{-\infty}^{+\infty}p_{\Xi_{\alpha}}\left(z\right)p_{\Lambda_{\alpha}}\left(z-x\right)dz,
\]
}
\par\end{center}{\small \par}
\end{spacing}

\noindent and therefore, by using Eq.~(\ref{eq:probability-distribution-decomposition}):

\begin{spacing}{0.8}
\begin{center}
{\small{}
\begin{align*}
P\left(\mathrm{len}\left(\mathcal{V}_{\alpha}\right)>0\right) & =\int_{0}^{+\infty}p_{\mathrm{len}\left(\mathcal{V}_{\alpha}\right)}\left(x\right)dx=\int_{-\infty}^{+\infty}p_{\Xi_{\alpha}}\left(x\right)F_{\Lambda_{\alpha}}\left(x\right)dx\\
\\
 & =\int_{-\infty}^{+\infty}p_{\Xi_{\alpha}^{c}}\left(x\right)F_{\Lambda_{\alpha}}\left(x\right)dx+\sum_{q\in\Gamma_{I_{\alpha},0}}\left[F_{\Xi_{\alpha}}\left(\theta_{q}\right)-\underset{x\rightarrow\theta_{q}^{-}}{\lim}F_{\Xi_{\alpha}}\left(x\right)\right]F_{\Lambda_{\alpha}}\left(\theta_{q}\right).
\end{align*}
}
\par\end{center}{\small \par}
\end{spacing}

\noindent To conclude, since the quantities $\mathrm{len}\left(\mathcal{V}_{\alpha}\right)$
are independent, we obtain that the probability to observe a given
firing-rate state $\boldsymbol{\nu}$ in the whole multistability
diagram of a single network realization is:
\begin{spacing}{0.8}
\begin{center}
{\small{}
\begin{equation}
P\left(\mathrm{vol}\left(\mathscr{V}\right)>0\right)=\prod_{\alpha=0}^{\mathfrak{P}-1}P\left(\mathrm{len}\left(\mathcal{V}_{\alpha}\right)>0\right).\label{eq:probability-regardless-of-the-stimuli}
\end{equation}
}
\par\end{center}{\small \par}
\end{spacing}

\subsection{The Special Case of Multi-Population Networks Composed of Statistically-Homogeneous
Populations \label{subsec:The-Special-Case-of-Multi-Population-Networks-Composed-of-Statistically-Homogeneous-Populations}}

In biological networks, heterogeneity is experimentally observed between
different types of synapses (e.g. excitatory vs inhibitory ones),
as well as within a given synaptic type \cite{Marder2006}. For this
reason, in this subsection we focused our attention on the study of
random networks composed of $\mathfrak{P}$ statistically-homogeneous
populations. As we explained in SubSec.~(\ref{subsec:Mean-Multistability-Diagram}),
by statistical homogeneity we mean that the synaptic weights are random
and therefore heterogeneous, but the probability distribution of $\mathscr{S}_{i}^{\left(\beta\right)}\overset{\mathrm{def}}{=}\sum_{k:\,\mathfrak{p}\left(k\right)=\beta}J_{i,k}$,
as well as the firing threshold $\theta_{i}$ and the external stimulus
$\mathfrak{I}_{i}$, are population-dependent. This model has been
used previously in neuroscience to study the dynamical consequences
of heterogeneous synaptic connections in multi-population networks
(see e.g. \cite{Sompolinsky1988,Hermann2012}). However, while previous
studies focused on the thermodynamic limit of the network model, here
we considered the case of arbitrary-size networks.

We called $N_{\alpha}$ the size of population $\alpha$, so that
$\sum_{\alpha=0}^{\mathfrak{P}-1}N_{\alpha}=N$. Moreover, we rearranged
the neurons so that the connection probabilities can be written in
the following block-matrix form:

\begin{spacing}{0.8}
\begin{center}
{\small{}
\begin{equation}
\mathcal{P}=\left[\begin{array}{cccc}
\mathscr{P}_{0,0} & \mathscr{P}_{0,1} & \cdots & \mathscr{P}_{0,\mathfrak{P}-1}\\
\mathscr{P}_{1,0} & \mathscr{P}_{1,1} & \cdots & \mathscr{P}_{1,\mathfrak{P}-1}\\
\vdots & \vdots & \ddots & \vdots\\
\mathscr{P}_{\mathfrak{P}-1,0} & \mathscr{P}_{\mathfrak{P}-1,1} & \cdots & \mathscr{P}_{\mathfrak{P}-1,\mathfrak{P}-1}
\end{array}\right],\quad\mathscr{P}_{\alpha,\beta}=\begin{cases}
P_{\alpha}^{\mathrm{aut}}\mathrm{Id}_{N_{\alpha}}+P_{\alpha,\alpha}\left(\mathbb{I}_{N_{\alpha}}-\mathrm{Id}_{N_{\alpha}}\right), & \,\mathrm{if}\;\alpha=\beta\\
\\
P_{\alpha,\beta}\mathbb{I}_{N_{\alpha},N_{\beta}}, & \,\mathrm{if}\;\alpha\neq\beta.
\end{cases}\label{eq:connection-probabilities}
\end{equation}
}
\par\end{center}{\small \par}
\end{spacing}

\noindent In Eq.~(\ref{eq:connection-probabilities}), $\mathscr{P}_{\alpha,\beta}$
is a $N_{\alpha}\times N_{\beta}$ matrix, while $P_{\alpha}^{\mathrm{aut}}$
represents the magnitude of the diagonal entries of the matrix $\mathscr{P}_{\alpha,\alpha}$,
namely the probability to observe a self-connection or autapse \cite{Yilmaz2016}.
$P_{\alpha,\beta}$ represents the probability to observe a synaptic
connection from a neuron in population $\beta$ to a (distinct) neuron
in population $\alpha$. Moreover, $\mathbb{I}_{N_{\alpha},N_{\beta}}$
is the $N_{\alpha}\times N_{\beta}$ all-ones matrix (here we used
the simplified notation $\mathbb{I}_{N_{\alpha}}\overset{\mathrm{def}}{=}\mathbb{I}_{N_{\alpha},N_{\alpha}}$),
while $\mathrm{Id}_{N_{\alpha}}$ is the $N_{\alpha}\times N_{\alpha}$
identity matrix. According to the homogeneity assumption, we also
supposed that the strength of the non-zero synaptic connections from
population $\beta$ to population $\alpha$ is generated from a population-dependent
probability distribution:
\begin{spacing}{0.8}
\begin{center}
{\small{}
\[
p_{W_{i,j}}=\begin{cases}
p_{\alpha}^{\left(\mathrm{aut}\right)} & \mathrm{for}\;\;i=j\\
\\
p_{\alpha,\beta} & \mathrm{otherwise}.
\end{cases}
\]
}
\par\end{center}{\small \par}
\end{spacing}

\noindent $\forall i,\;j$ belonging to populations $\alpha,\;\beta$,
respectively. For every excitatory population the support of the distribution
$p_{W_{i,j}}$ must be non-negative, while for every inhibitory population
it must be non-positive. We also supposed that all the neurons in
population $\alpha$ have the same firing threshold $\theta_{i}=\vartheta_{\alpha}$
and share the same stimulus $\mathfrak{I}_{i}=I_{\alpha}$. For example,
if each population receives a distinct external stimulus, then $\Gamma_{I_{\alpha}}=\left\{ n_{\alpha-1},n_{\alpha-1}+1,\cdots,n_{\alpha}-1\right\} $,
where $n_{\alpha-1}\overset{\mathrm{def}}{=}{\displaystyle \sum_{\beta=0}^{\alpha-1}}N_{\beta}$
and $n_{-1}\overset{\mathrm{def}}{=}0$. However, generally, there
may exist distinct populations that share the same stimulus.

Now consider the following $\mathscr{N}\times\mathscr{N}$ block matrix:
\begin{spacing}{0.8}
\begin{center}
{\small{}
\[
\mathcal{B}\overset{\mathrm{def}}{=}\left[\begin{array}{cccc}
\mathscr{B}_{0,0} & \mathscr{B}_{0,1} & \cdots & \mathscr{B}_{0,\mathfrak{Y}-1}\\
\mathscr{B}_{1,0} & \mathscr{B}_{1,1} & \cdots & \mathscr{B}_{1,\mathfrak{Y}-1}\\
\vdots & \vdots & \ddots & \vdots\\
\mathscr{B}_{\mathfrak{X}-1,0} & \mathscr{B}_{\mathfrak{X}-1,1} & \cdots & \mathscr{B}_{\mathfrak{X}-1,\mathfrak{Y}-1}
\end{array}\right],
\]
}
\par\end{center}{\small \par}
\end{spacing}

\noindent with homogeneous $X_{\lambda}\times Y_{\mu}$ blocks $\mathscr{B}_{\lambda,\mu}=B_{\lambda,\mu}\mathbb{I}_{X_{\lambda},Y_{\mu}}$
(where $\sum_{\lambda=0}^{\mathfrak{X}-1}X_{\lambda}=\sum_{\mu=0}^{\mathfrak{Y}-1}Y_{\mu}=\mathscr{N}$,
while $B_{\lambda,\mu}$ are free parameters). We found that:
\begin{spacing}{0.8}
\begin{center}
{\small{}
\begin{align}
 & \mathrm{per}\left(\mathcal{B}\right)=\mathcal{C}\sum_{\boldsymbol{s}\in\mathcal{S}}\mathcal{T}_{\boldsymbol{s}}\mathcal{U}_{\boldsymbol{s}}\nonumber \\
\nonumber \\
 & \mathcal{S}\overset{\mathrm{def}}{=}\left\{ \boldsymbol{s}=\left(s_{0,0},\cdots,s_{\mathfrak{X}-1,\mathfrak{Y}-1}\right)\in\mathbb{N}^{\mathfrak{X}\mathfrak{Y}}:\;\sum_{\mu=0}^{\mathfrak{Y}-1}s_{\lambda,\mu}=X_{\lambda}\;\forall\lambda,\;\sum_{\lambda=0}^{\mathfrak{X}-1}s_{\lambda,\mu}=Y_{\mu}\;\forall\mu\right\} ,\quad\mathbb{N}\overset{\mathrm{def}}{=}\left\{ 0,1,2,\cdots\right\} \label{eq:permanent-homogeneous-block-matrix}\\
\nonumber \\
 & \mathcal{C}\overset{\mathrm{def}}{=}\prod_{\lambda=0}^{\mathfrak{X}-1}\left(X_{\lambda}!\right),\quad\mathcal{T}_{\boldsymbol{s}}\overset{\mathrm{def}}{=}\prod_{\mu=0}^{\mathfrak{Y}-1}\binom{Y_{\mu}}{s_{0,\mu},\cdots,s_{\mathfrak{X}-1,\mu}},\quad\mathcal{U}_{\boldsymbol{s}}\overset{\mathrm{def}}{=}\prod_{\lambda=0}^{\mathfrak{X}-1}\prod_{\mu=0}^{\mathfrak{Y}-1}B_{\lambda,\mu}^{s_{\lambda,\mu}},\nonumber 
\end{align}
}
\par\end{center}{\small \par}
\end{spacing}

\noindent with multinomial coefficients:
\begin{spacing}{0.8}
\noindent \begin{center}
{\small{}
\[
\binom{Y_{\mu}}{s_{0,\mu},\cdots,s_{\mathfrak{X}-1,\mu}}\overset{\mathrm{def}}{=}\frac{Y_{\mu}!}{s_{0,\mu}!\cdots s_{\mathfrak{X}-1,\mu}!}.
\]
}
\par\end{center}{\small \par}
\end{spacing}

\noindent As a consequence of the statistical homogeneity of the multi-population
network considered in this subsection, the matrices in Eqs.~(\ref{eq:continuous-components})
and (\ref{eq:cumulative-distribution-functions}) are composed of
homogeneous block submatrices. For this reason, in the specific case
of this multi-population network, the permanents in Eqs.~(\ref{eq:continuous-components})
and (\ref{eq:cumulative-distribution-functions}) can be calculated
by means of Eq.~(\ref{eq:permanent-homogeneous-block-matrix}). Note
that, for a given $\alpha$, the parameter $\mathfrak{X}$ represents
the number of distinct populations that share the current $I_{\alpha}$
(for example, $\mathfrak{X}=1$ if each population receives a distinct
external stimulus), while $\mathfrak{Y}=1,2$ is the number of block
columns (for example, $\mathfrak{Y}=1$ when calculating $F_{\Lambda_{\alpha}}\left(x\right)$,
see Eq.~(\ref{eq:cumulative-distribution-functions}), and $\mathfrak{Y}=2$
when calculating $p_{\Lambda_{\alpha}^{c}}\left(x\right)$, see Eq.~(\ref{eq:continuous-components})).
$X_{\lambda}$ corresponds to the number of neurons with index $i\in\Gamma_{I_{\alpha},u}$
that belong to the population $\lambda$, while $Y_{\mu}$ represents
the number of columns of the $\mu$th block matrix (for example, when
calculating $p_{\Lambda_{\alpha}^{c}}\left(x\right)$, we set $Y_{0}=1$
and $Y_{1}=\gamma_{\alpha,1}-1$, which correspond to the number of
columns of the submatrices $\boldsymbol{a}_{\alpha,1}\left(\boldsymbol{\theta}-\boldsymbol{x}\right)$
and $\boldsymbol{F}_{\alpha,1}^{\left(\gamma_{\alpha,1}-1\right)}\left(x\right)$
respectively, see Eq.~(\ref{eq:continuous-components})). Moreover,
$\mathscr{N}$ corresponds to $\gamma_{\alpha,u}$, while the parameters
$B_{\lambda,\mu}$ represent the entries of the matrices in Eqs.~(\ref{eq:continuous-components})
and (\ref{eq:cumulative-distribution-functions}) (for example, $\mathscr{N}=\gamma_{\alpha,1}$,
$B_{\lambda,0}=p_{\mathcal{I}_{i}}\left(x\right)$ and $B_{\lambda,1}=F_{\mathcal{I}_{i}}\left(x\right)$
for $i$ in the population $\lambda$, when calculating $p_{\Lambda_{\alpha}^{c}}\left(x\right)$).

For the sake of clarity, we implemented Eq.~(\ref{eq:permanent-homogeneous-block-matrix})
in the supplemental Python script ``Permanent.py''. Since the permanents
in Eqs.~(\ref{eq:continuous-components}) and (\ref{eq:cumulative-distribution-functions})
can be obtained from Eq.~(\ref{eq:permanent-homogeneous-block-matrix})
for $\mathfrak{Y}=1$ and $\mathfrak{Y}=2$, in the script we specifically
implemented these two cases. The computation of $\mathrm{per}\left(\mathcal{B}\right)$
by means of Eq.~(\ref{eq:permanent-homogeneous-block-matrix}) generally
proved much faster than the BBFG algorithm, see Sec.~(\ref{sec:Results}).
However, it is important to note that while the BBFG algorithm can
be applied to neural networks with any topology, Eq.~(\ref{eq:permanent-homogeneous-block-matrix})
is specific for multi-population networks composed of statistically-homogeneous
populations.

\subsection{Large-Network Limit \label{subsec:Large-Network-Limit}}

In computational neuroscience, statistically-homogeneous multi-population
networks represent an important class of network models, since their
large-size limit is typically well-defined and serves as a basis for
understanding the asymptotic behavior of neural systems \cite{Sompolinsky1988,Cessac1995,Faugeras2009,Hermann2012,Cabana2013}.
In this subsection, we derived the large-size limit of the class of
statistically-homogeneous multi-population networks with quenched
disorder that we introduced in SubSec.~(\ref{subsec:The-Special-Case-of-Multi-Population-Networks-Composed-of-Statistically-Homogeneous-Populations}).
In particular, we focused on the case when each neural population
receives a distinct external stimulus current, and we also supposed
that the contribution of self-connections to the statistics of the
firing rates is negligible in the large-network limit. The consequences
of the relaxation of these two assumptions will be discussed at the
end of this subsection.

The derivation of the asymptotic form of the stationary-state statistics
required the introduction of a proper normalization of the sum $S_{i}=\sum_{j=0}^{N-1}J_{i,j}\nu_{j}$
in Eq.~(\ref{eq:synchronous-network-equations}), in order to prevent
the divergence of the mean and the variance of $S_{i}$ in the thermodynamic
limit. To this purpose, we chose the mean and the variance of the
random variables $W_{i,j}$ as follows:
\begin{spacing}{0.8}
\begin{center}
{\small{}
\begin{align}
 & m_{\alpha,\beta}\overset{\mathrm{def}}{=}\left\langle W_{i,j}\right\rangle =\frac{\mu_{\alpha,\beta}}{N_{\beta}}\nonumber \\
\label{eq:normalization}\\
 & s_{\alpha,\beta}^{2}\overset{\mathrm{def}}{=}\mathrm{Var}\left(W_{i,j}\right)=\frac{\sigma_{\alpha,\beta}^{2}}{N_{\beta}}+\left(\frac{\mu_{\alpha,\beta}}{N_{\beta}}\right)^{2}\left(P_{\alpha,\beta}-1\right),\nonumber 
\end{align}
}
\par\end{center}{\small \par}
\end{spacing}

\noindent given parameters $\mu_{\alpha,\beta}$, $\sigma_{\alpha,\beta}$,
$N_{\beta}$ and $P_{\alpha,\beta}$ such that $\mu_{\alpha,\beta}\in\mathbb{R}$
and $\mathrm{Var}\left(W_{i,j}\right)\in\mathbb{R}_{\geq0}$, for
every $i,\;j$ (with $i\neq j$) in the populations $\alpha,\;\beta$,
respectively. Eq.~(\ref{eq:normalization}) implies that:
\begin{spacing}{0.8}
\begin{center}
{\small{}
\begin{align*}
 & \left\langle J_{i,j}\right\rangle =\left\langle T_{i,j}\right\rangle \left\langle W_{i,j}\right\rangle =\frac{P_{\alpha,\beta}}{N_{\beta}}\mu_{\alpha,\beta}\\
\\
 & \mathrm{Var}\left(J_{i,j}\right)=\left\langle T_{i,j}^{2}\right\rangle \left\langle W_{i,j}^{2}\right\rangle -\left\langle T_{i,j}\right\rangle ^{2}\left\langle W_{i,j}\right\rangle ^{2}=\frac{P_{\alpha,\beta}}{N_{\beta}}\sigma_{\alpha,\beta}^{2},
\end{align*}
}
\par\end{center}{\small \par}
\end{spacing}

\noindent and therefore:
\begin{spacing}{0.8}
\begin{center}
{\small{}
\begin{align*}
 & \mu_{\alpha}\overset{\mathrm{def}}{=}\left\langle \sum_{j=0}^{N-1}J_{i,j}\nu_{j}\right\rangle \approx\sum_{\beta=0}^{\mathfrak{P}-1}\frac{\gamma_{\beta,1}P_{\alpha,\beta}}{N_{\beta}}\mu_{\alpha,\beta}\\
\\
 & \sigma_{\alpha}^{2}\overset{\mathrm{def}}{=}\mathrm{Var}\left(\sum_{j=0}^{N-1}J_{i,j}\nu_{j}\right)\approx\sum_{\beta=0}^{\mathfrak{P}-1}\frac{\gamma_{\beta,1}P_{\alpha,\beta}}{N_{\beta}}\sigma_{\alpha,\beta}^{2},
\end{align*}
}
\par\end{center}{\small \par}
\end{spacing}

\noindent having neglected the contribution of the autapses. Therefore
the mean and the variance of $S_{i}$ are finite for every state $\boldsymbol{\nu}$
in the thermodynamic limit, as desired. Now, consider any of the firing-rate
states $\boldsymbol{\nu}$ composed of $\gamma_{\alpha,1}$ active
neurons in the population $\alpha$ ($\forall\alpha\in\left\{ 0,\cdots,\mathfrak{P}-1\right\} $).
For the central limit theorem, given any distribution (not necessarily
normal) of $W_{i,j}$ that satisfies Eq.~(\ref{eq:normalization}),
we get:
\begin{spacing}{0.8}
\begin{center}
{\small{}
\[
\sqrt{\gamma_{\beta,1}}\left(\frac{1}{\gamma_{\beta,1}}\sum_{j=n_{\beta-1}}^{n_{\beta}-1}J_{i,j}\nu_{j}-\frac{P_{\alpha,\beta}}{N_{\beta}}\mu_{\alpha,\beta}\right)\overset{d}{\rightarrow}\mathcal{N}\left(0,\frac{P_{\alpha,\beta}}{N_{\beta}}\sigma_{\alpha,\beta}^{2}\right)\;\forall\beta,
\]
}
\par\end{center}{\small \par}
\end{spacing}

\noindent in the limit $\gamma_{\beta,1}\rightarrow\infty$ (see SubSec.~(\ref{subsec:The-Special-Case-of-Multi-Population-Networks-Composed-of-Statistically-Homogeneous-Populations})
for the definition of the parameter $n_{\beta}$). In turn, this implies
that:
\begin{spacing}{0.8}
\noindent \begin{center}
{\small{}
\[
\mathcal{I}_{i}=\theta_{i}-\sum_{\beta=0}^{\mathfrak{P}-1}\sum_{j=n_{\beta-1}}^{n_{\beta}-1}J_{i,j}\nu_{j}\overset{d}{\rightarrow}\mathcal{N}\left(\vartheta_{\alpha}-\mu_{\alpha},\sigma_{\alpha}^{2}\right).
\]
}
\par\end{center}{\small \par}
\end{spacing}

Since the random variables $\mathcal{I}_{i}$ are independent and
identically distributed $\forall i\in\Gamma_{I_{\alpha}}$ and $\alpha$
fixed, according to the \textit{Fisher-Tippett-Gnedenko theorem} \cite{Fisher1928,Gnedenko1943,Gumbel1958},
the distribution of the variables $\Lambda_{\alpha}$ and $\Xi_{\alpha}$
converges to the \textit{Gumbel distribution} in the limit $\gamma_{\alpha,u}\rightarrow\infty$.
In other words, given $X\in\left\{ \Lambda,\Xi\right\} $, and by
defining, according to \cite{Vivo2015}:
\begin{spacing}{0.8}
\begin{center}
{\small{}
\begin{align}
 & z_{X_{\alpha}}\left(x\right)\overset{\mathrm{def}}{=}\frac{x-\mathfrak{a}_{X_{\alpha}}}{\mathfrak{b}_{X_{\alpha}}}\nonumber \\
 & \mathfrak{a}_{\Lambda_{\alpha}}\overset{\mathrm{def}}{=}\vartheta_{\alpha}-\mu_{\alpha}+\sigma_{\alpha}\Phi^{-1}\left(1-\frac{1}{\mathfrak{n}_{\Lambda_{\alpha}}}\right),\quad\mathfrak{a}_{\Xi_{\alpha}}\overset{\mathrm{def}}{=}\vartheta_{\alpha}-\mu_{\alpha}-\sigma_{\alpha}\Phi^{-1}\left(1-\frac{1}{\mathfrak{n}_{\Xi_{\alpha}}}\right)\nonumber \\
 & \mathfrak{b}_{X_{\alpha}}\overset{\mathrm{def}}{=}\sigma_{\alpha}\left[\Phi^{-1}\left(1-\frac{1}{\mathfrak{n}_{X_{\alpha}}e}\right)-\Phi^{-1}\left(1-\frac{1}{\mathfrak{n}_{X_{\alpha}}}\right)\right]\nonumber \\
\nonumber \\
 & \mathfrak{n}_{\Lambda_{\alpha}}\overset{\mathrm{def}}{=}\gamma_{\alpha,1},\quad\mathfrak{n}_{\Xi_{\alpha}}\overset{\mathrm{def}}{=}\gamma_{\alpha,0},\label{eq:Gumbel-parameters}
\end{align}
}
\par\end{center}{\small \par}
\end{spacing}

\noindent then in the limit $\mathfrak{n}_{X_{\alpha}}\rightarrow\infty$
we get:
\begin{spacing}{0.8}
\begin{center}
{\small{}
\begin{align}
 & p_{\Lambda_{\alpha}}\left(x\right)\rightarrow g\left(z_{\Lambda_{\alpha}}\left(x\right)\right),\quad F_{\Lambda_{\alpha}}\left(x\right)\rightarrow G\left(z_{\Lambda_{\alpha}}\left(x\right)\right)\nonumber \\
\label{eq:Gumbel-pdf-and-cdf}\\
 & p_{\Xi_{\alpha}}\left(x\right)\rightarrow g\left(-z_{\Xi_{\alpha}}\left(x\right)\right),\quad F_{\Xi_{\alpha}}\left(x\right)\rightarrow1-G\left(-z_{\Xi_{\alpha}}\left(x\right)\right),\nonumber 
\end{align}
}
\par\end{center}{\small \par}
\end{spacing}

\noindent where $g\left(\cdot\right)$ and $G\left(\cdot\right)$
are the Gumbel probability density and its cumulative distribution
function, respectively:
\begin{spacing}{0.8}
\begin{center}
{\small{}
\[
g\left(x\right)=\frac{1}{\mathfrak{b}_{X_{\alpha}}}e^{-\left(x+e^{-x}\right)},\quad G\left(x\right)=e^{-e^{-x}}.
\]
}
\par\end{center}{\small \par}
\end{spacing}

\noindent In Eq.~(\ref{eq:Gumbel-parameters}), $\Phi\left(\cdot\right)$
represents the cumulative distribution function of the standard normal
probability density. $\Phi^{-1}\left(\cdot\right)$ is the \textit{probit
function}, which can be expressed in terms of the inverse of the error
function $\mathrm{erf}\left(\cdot\right)$ as $\Phi^{-1}\left(x\right)=\sqrt{2}\mathrm{erf}^{-1}\left(2x-1\right)$.
By using an asymptotic expansion of $\mathrm{erf}^{-1}\left(\cdot\right)$,
we get:
\begin{spacing}{0.8}
\begin{center}
{\small{}
\begin{align*}
\Phi^{-1}\left(1-\frac{1}{\mathfrak{n}_{X_{\alpha}}}\right)\approx & \sqrt{\ln\left(\frac{\mathfrak{n}_{X_{\alpha}}^{2}}{2\pi}\right)-\ln\left(\ln\left(\frac{\mathfrak{n}_{X_{\alpha}}^{2}}{2\pi}\right)\right)}\\
\\
\Phi^{-1}\left(1-\frac{1}{\mathfrak{n}_{X_{\alpha}}e}\right)\approx & \sqrt{2+\ln\left(\frac{\mathfrak{n}_{X_{\alpha}}^{2}}{2\pi}\right)-\ln\left(2+\ln\left(\frac{\mathfrak{n}_{X_{\alpha}}^{2}}{2\pi}\right)\right)}.
\end{align*}
}
\par\end{center}{\small \par}
\end{spacing}

\noindent Moreover, it is possible to prove that:
\begin{spacing}{0.8}
\begin{center}
{\small{}
\begin{equation}
\left\langle \Lambda_{\alpha}\right\rangle =\mathfrak{a}_{\Lambda_{\alpha}}+\mathfrak{b}_{\Lambda_{\alpha}}\gamma,\quad\left\langle \Xi_{\alpha}\right\rangle =\mathfrak{a}_{\Xi_{\alpha}}-\mathfrak{b}_{\Xi_{\alpha}}\gamma,\label{eq:mean-of-the-distributions}
\end{equation}
}
\par\end{center}{\small \par}
\end{spacing}

\noindent where $\gamma$ is the Euler-Mascheroni constant. Note that
Eq.~(\ref{eq:mean-of-the-distributions}) can be used for plotting
the mean multistability diagram of the network, while Eqs.~(\ref{eq:probability-for-fixed-stimuli}),
(\ref{eq:Gumbel-parameters}) and (\ref{eq:Gumbel-pdf-and-cdf}) provide
an analytical expression of the occurrence probability of the stationary
states for a given combination of stimuli. Unfortunately, we are not
aware of any exact formula of the occurrence probability of the stationary
states regardless of the stimuli (see SubSec.~\ref{subsec:Occurrence-Probability-of-the-Stationary-States-Regardless-of-the-Stimuli}).
For this reason, the latter should be calculated numerically or through
analytical approximations, from Eq.~(\ref{eq:probability-regardless-of-the-stimuli}).

Now we discuss the two assumptions that we made in the derivation
of our results, namely distinct external stimuli to each neural population,
and a negligible contribution of the autapses to the statistics of
the firing rates. The relaxation of the first assumption implies the
calculation of the minimum and maximum of non-identically distributed
random variables. For example, in the case when an external stimulus
is shared by two distinct populations, the variable $\mathcal{I}_{i}$
has two distinct probability distributions, depending on the population
the neuron $i$ belongs to. However, the Fisher-Tippett-Gnedenko theorem
is valid only for identically distributed variables, and a straightforward
generalization of the theorem to the case of non-identically distributed
variables is not available (see SubSec.~(\ref{subsec:Limitations-of-Our-Approach})).
Note, however, that this limitation applies only to the asymptotic
expansion discussed in the present subsection. The exact (i.e. non-asymptotic)
theory discussed in SubSec.~(\ref{subsec:Statistical-Properties-of-the-Network-Model})
is not affected by this limitation, and is valid also when a stimulus
is shared by several populations.

The second assumption in our derivation was the negligible contribution
of the autapses to the statistics of the firing rates in the large-network
limit. This assumption can be relaxed for example by supposing that
the autapses are not scaled, so that the random variable $S_{i}$
is strongly affected by the autaptic weight $J_{i,i}$ when $\nu_{i}=1$.
In this case, the central limit theorem does not apply anymore to
the whole sum $S_{i}=\sum_{j=0}^{N-1}J_{i,j}\nu_{j}$. In other words,
in the case when the autapses are not normally distributed, the sum
$S_{i}$ is not normally distributed either, therefore the distribution
of the variables $\Lambda_{\alpha}$ and $\Xi_{\alpha}$ may not be
necessarily the Gumbel law. This case can be studied analytically,
if desired, but we omitted it for the sake of brevity.

\subsection{Numerical Simulations \label{subsec:Numerical-Simulations}}

To further validate our results, in Sec.~(\ref{sec:Results}) we
compared our semi-analytical formulas with numerical Monte Carlo simulations,
that we implemented in the ``Numerical Simulations'' section of
the supplemental Python script ``Statistics.py''. During these numerical
simulations, we ran a large number of network realizations ($5,000$
for the results shown in Figs.~(\ref{fig:cumulative-distribution-functions})-(\ref{fig:occurrence-probability-firing-rate-states}),
and $100,000$ for those in Fig.~(\ref{fig:large-size-limit})),
and at each of them we generated a new (quenched) connectivity matrix
$J$, according to Eq.~(\ref{eq:synaptic-weights-distribution}).
Then, for each $J$, we derived the corresponding bifurcation points
$\Lambda_{\alpha}$ and $\Xi_{\alpha}$ and the hypervolumes $\mathscr{V}$,
by applying the algorithm ``Multistability\_Diagram.py'' that we
introduced in \cite{Fasoli2018b}.

The cumulative distribution function of the bifurcation points was
then computed by means of a cumulative sum of the probability histograms
of $\Lambda_{\alpha}$ and $\Xi_{\alpha}$. This provided a numerical
approximation of the functions $F_{\Lambda_{\alpha}}\left(x\right)$
and $F_{\Xi_{\alpha}}\left(x\right)$, that we derived semi-analytically
in Eq.~(\ref{eq:cumulative-distribution-functions}).

The mean multistability diagram was calculated by averaging the bifurcation
points $\Lambda_{\alpha}$ and $\Xi_{\alpha}$ over the network realizations.
This provided a numerical approximation of the quantities $\left\langle \Lambda_{\alpha}\right\rangle $
and $\left\langle \Xi_{\alpha}\right\rangle $, that we derived semi-analytically
in Eqs.~(\ref{eq:mean_method_1}) and (\ref{eq:mean_method_2}).

The probability that a given firing-rate state $\boldsymbol{\nu}$
is stationary for a fixed combination of stimuli $\widehat{\boldsymbol{I}}$
was calculated by counting, during the Monte Carlo simulations, the
number of times $\widehat{\boldsymbol{I}}\in\mathscr{V}$. By dividing
this number by the total number of realizations, we obtained a numerical
estimation of the probability $P\left(\widehat{\boldsymbol{I}}\in\mathscr{V}\right)$
(see Eq.~(\ref{eq:probability-for-fixed-stimuli}) for its semi-analytical
expression). This calculation was then repeated for each of the $2^{N}$
firing-rate states $\boldsymbol{\nu}$. Alternatively, this probability
can be calculated by counting the relative number of times $\boldsymbol{\nu}\left(t_{0}+1\right)=\boldsymbol{\nu}\left(t_{0}\right)$
(stationarity condition), where the firing-rate state $\boldsymbol{\nu}\left(t_{0}\right)$
is each of the $2^{N}$ initial conditions of the network model, while
the state $\boldsymbol{\nu}\left(t_{0}+1\right)$ is calculated iteratively
from it by means of Eqs.~(\ref{eq:synchronous-network-equations})
or (\ref{eq:asynchronous-network-equations}). We implemented both
methods in the Python script ``Statistics.py'', and the reader can
easily check that they provide identical numerical estimations of
$P\left(\widehat{\boldsymbol{I}}\in\mathscr{V}\right)$.

The probability that the state $\boldsymbol{\nu}$ is stationary,
regardless of the specific combination of stimuli to the network,
was derived numerically by counting the relative number of times $\mathrm{vol}\left(\mathscr{V}\right)=\prod_{\alpha=0}^{\mathfrak{P}-1}\mathrm{len}\left(\mathcal{V}_{\alpha}\right)>0$,
for each of the $2^{N}$ firing-rate states $\boldsymbol{\nu}$. This
provided a numerical estimation of the probability $P\left(\mathrm{vol}\left(\mathscr{V}\right)>0\right)$,
that we derived semi-analytically in Eq.~(\ref{eq:probability-regardless-of-the-stimuli}).

To conclude, the probability distribution of the bifurcation points
in the large-size limit was calculated numerically through a kernel
density estimation, in the specific case of statistically-homogeneous
multi-population networks. The density estimator was applied to the
samples of the random variables $\Lambda_{\alpha}$ and $\Xi_{\alpha}$,
which were generated during the Monte Carlo simulations according
to Eq.~(\ref{eq:stationary-states-stimulus-range}), for a given
firing-rate state $\boldsymbol{\nu}$.

\section{Results \label{sec:Results}}

In this section we reported the comparison between, on one hand, the
semi-analytical formulas of the mean multistability diagram, of the
occurrence probability of the stationary firing-rate states, and of
the probability distribution of the bifurcation points in both the
small and large-size limits (see SubSecs.~(\ref{subsec:Statistical-Properties-of-the-Network-Model})
- (\ref{subsec:Large-Network-Limit})), and, on the other hand, the
corresponding numerical counterparts (SubSec.~(\ref{subsec:Numerical-Simulations})).

For illustrative purposes, we considered the case $\mathfrak{P}=2$,
so that the multistability diagram can be visualized on a plane. In
particular, we supposed that the network is composed of excitatory
($E$) and inhibitory ($I$) neurons. For this reason, it is convenient
to change the notation slightly, and to consider $\alpha\in\left\{ E,I\right\} $
rather than $\alpha\in\left\{ 0,1\right\} $ (so that the multistability
diagram will be plotted on the $I_{E}-I_{I}$ plane). Since the total
number of firing-rate states of the network increases as $2^{N}$
with the network size, in this section we applied the Python script
``Statistics.py'' to a small-sized network ($N=4$), in order to
ensure the clarity of the figures. It is important to note that, in
the derivation of the results in SubSecs.~(\ref{subsec:Statistical-Properties-of-the-Network-Model})
and (\ref{subsec:The-Special-Case-of-Multi-Population-Networks-Composed-of-Statistically-Homogeneous-Populations}),
we did not resort to any mathematical approximation, and that our
semi-analytical formulas are exact for every size $N$. For this reason,
if desired, our script can be applied to networks with size $N\gg4$,
depending on the computational power available.

\begin{figure}
\begin{centering}
\includegraphics[scale=0.24]{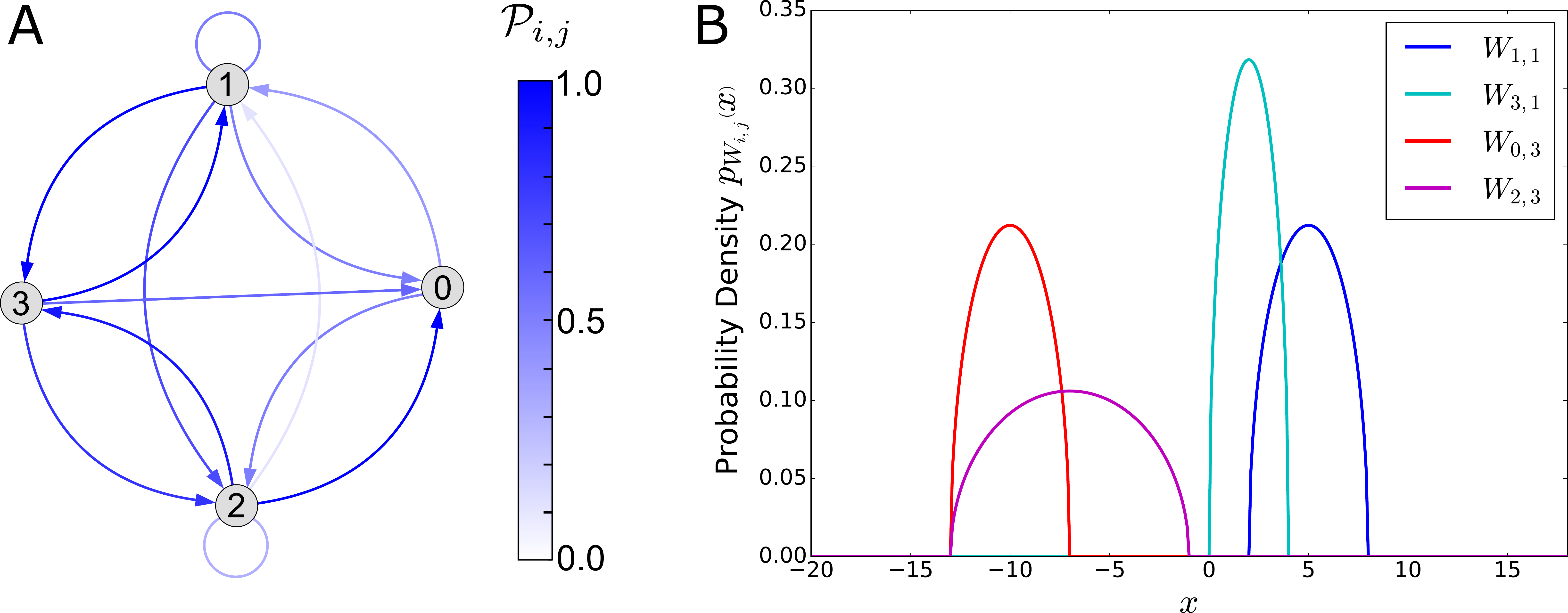}
\par\end{centering}
\caption{\label{fig:synaptic-weights} \small\textbf{ Probability distribution
of the synaptic weights.} This figure shows the probability distribution
of the synaptic weights $J_{i,j}$ (see Eq.~(\ref{eq:synaptic-weights})),
in the specific case of the small-size network model described in
Sec.~(\ref{sec:Results}). A) Graph of the connection probability
matrix $\mathcal{P}$ reported in Tab.~(\ref{tab:Parameters-0}).
An arrow from the vertex $j$ to the vertex $i$ represents a connection
probability $\mathcal{P}_{i,j}>0$. B) Wigner semicircle distribution
of some variables $W_{i,j}$, according to Eq.~(\ref{eq:Wigner-semicircle-distribution})
and the values of the parameters $\mathfrak{C}$ and $\mathfrak{R}$
reported in Tab.~(\ref{tab:Parameters-0}).}
\end{figure}

In the network that we tested, we supposed that the neurons with indexes
$i\in\left\{ 0,1\right\} $ are excitatory and receive an external
stimulus $I_{E}$, while the neurons with indexes $i\in\left\{ 2,3\right\} $
are inhibitory and receive a stimulus $I_{I}$. Moreover, we supposed
that the independent random variables $W_{i,j}$ are distributed according
to the following Wigner semicircle distribution:
\begin{spacing}{0.8}
\begin{center}
{\small{}
\begin{equation}
p_{W_{i,j}}\left(x\right)=\begin{cases}
\frac{2}{\pi\mathfrak{R}_{i,j}^{2}}\sqrt{\mathfrak{R}_{i,j}^{2}-\left(x-\mathfrak{C}_{i,j}\right)^{2}}, & \mathrm{if}\;\;\left|x-\mathfrak{C}_{i,j}\right|\leq\mathfrak{R}_{i,j}\\
\\
0, & \mathrm{otherwise},
\end{cases}\label{eq:Wigner-semicircle-distribution}
\end{equation}
}
\par\end{center}{\small \par}
\end{spacing}

\noindent centered at $x=\mathfrak{C}_{i,j}$ and with radius $\mathfrak{R}_{i,j}$.
In Tab.~(\ref{tab:Parameters-0}) we reported the values of the parameters
$\mathcal{P}$, $\boldsymbol{\theta}=\left[\begin{array}{cccc}
\theta_{0}, & \theta_{1}, & \theta_{2}, & \theta_{3}\end{array}\right]$, $\mathfrak{C}$ and $\mathfrak{R}$ that we chose for this network.
\begin{table}
\begin{centering}
\textbf{\small{}}%
\begin{tabular}{|lllll|}
\hline 
 &  &  &  & \tabularnewline
 & {\small{}$\mathcal{P}=\left[\begin{array}{cccc}
0 & 0.5 & 1 & 0.6\\
0.4 & 0.5 & 0.1 & 1\\
0.5 & 0.7 & 0.3 & 0.8\\
0 & 1 & 0.9 & 0
\end{array}\right],$} &  & {\small{}$\boldsymbol{\theta}=\left[\begin{array}{c}
0\\
1\\
1\\
2
\end{array}\right]$} & \tabularnewline
 &  &  &  & \tabularnewline
 & {\small{}$\mathfrak{C}=\left[\begin{array}{cccc}
\times & 4 & -3 & -10\\
6 & 5 & -2 & -4\\
3 & 4 & -6 & -7\\
\times & 2 & -5 & \times
\end{array}\right],$} &  & {\small{}$\mathfrak{R}=\left[\begin{array}{cccc}
\times & 4 & 2 & 3\\
5 & 3 & 2 & 3\\
3 & 4 & 5 & 6\\
\times & 2 & 4 & \times
\end{array}\right]$} & \tabularnewline
 &  &  &  & \tabularnewline
\hline 
\end{tabular}
\par\end{centering}{\small \par}
\caption{\label{tab:Parameters-0} \textbf{An example of network parameters}.
This table contains the values of the parameters of the small-size
network that we studied in Sec.~(\ref{sec:Results}) (see Eq.~(\ref{eq:Wigner-semicircle-distribution})
and Fig.~(\ref{fig:synaptic-weights})). The symbol $\times$ in
the matrices $\mathfrak{C}$ and $\mathfrak{R}$ means that the statistics
of the stationary states and of the bifurcation points are not affected
by those parameters, since the corresponding synaptic connections
are absent ($\mathcal{P}_{i,j}=0$).}
\end{table}
 In panel A of Fig.~(\ref{fig:synaptic-weights}) we showed the graph
of the connection probability matrix $\mathcal{P}$, while in panel
B we plotted some examples of the Wigner probability distributions
$p_{W_{i,j}}$. 
 Note that, for our choice of the parameters, the support of the Wigner
distribution, namely the range $\left[\mathfrak{C}_{i,j}-\mathfrak{R}_{i,j},\mathfrak{C}_{i,j}+\mathfrak{R}_{i,j}\right]$,
is a subset of $\mathbb{R}_{\geq0}$ (respectively $\mathbb{R}_{\leq0}$)
for excitatory (respectively inhibitory) neurons. It follows that
the connectivity matrix $J$ of the model satisfies the Dale's principle
\cite{Strata1999}, as required for biologically realistic networks
(note, however, that our algorithm can be applied also to networks
that do not satisfy the principle, if desired).

\begin{figure}
\begin{centering}
\includegraphics[scale=0.26]{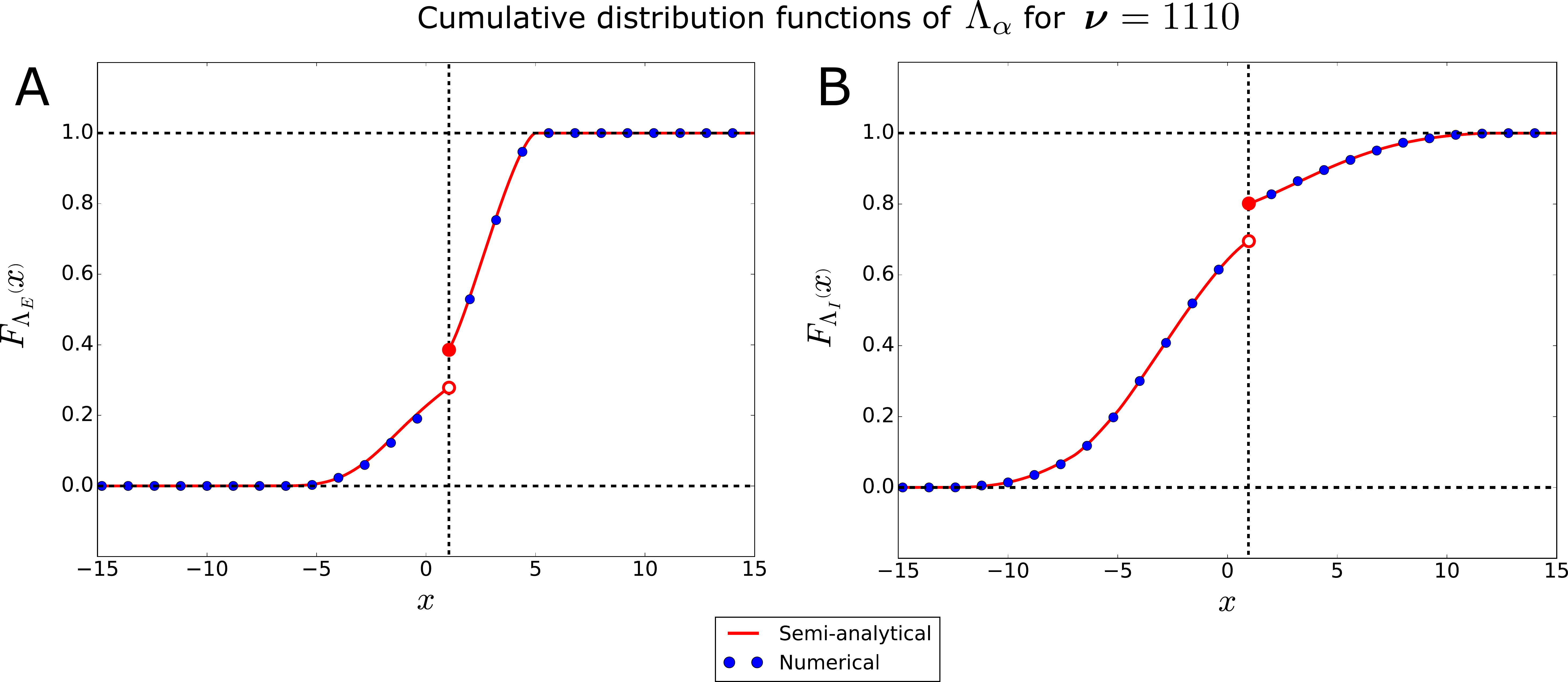}
\par\end{centering}
\caption{\label{fig:cumulative-distribution-functions} \small\textbf{ Examples
of cumulative distribution functions of the bifurcation points.} This
figure reports the cumulative distribution functions of the bifurcation
points $\Lambda_{E}$ and $\Lambda_{I}$ of the firing-rate state
$1110$ (panels A and B, respectively), in the case of the small-size
network described in SubSec.~(\ref{sec:Results}). The red curves
represent the semi-analytical functions, calculated through Eq.~(\ref{eq:cumulative-distribution-functions}),
while the blue dots represent the numerical functions, computed over
$5,000$ network realizations as described in SubSec.~(\ref{subsec:Numerical-Simulations}).
Similar results can be derived for all the other bifurcation points
of the network, if desired.}
\end{figure}

In Fig.~(\ref{fig:cumulative-distribution-functions}) we plotted
the cumulative distribution functions of the bifurcation points $\Lambda_{E}$
and $\Lambda_{I}$ of, e.g., the firing-rate state $1110$. 
 The figure shows a very good agreement between the semi-analytical
functions (red curves), calculated through Eq.~(\ref{eq:cumulative-distribution-functions}),
and the numerical functions (blue dots), computed over $5,000$ network
realizations as described in SubSec.~(\ref{subsec:Numerical-Simulations}).
Note that, generally, the cumulative distribution functions are not
continuous (see also SubSec.~(\ref{subsec:Probability-Distribution-of-the-Bifurcation-Points})),
and that jump discontinuities may occur at the firing thresholds ($\theta=1$,
in this example).

\begin{figure}
\begin{centering}
\includegraphics[scale=0.35]{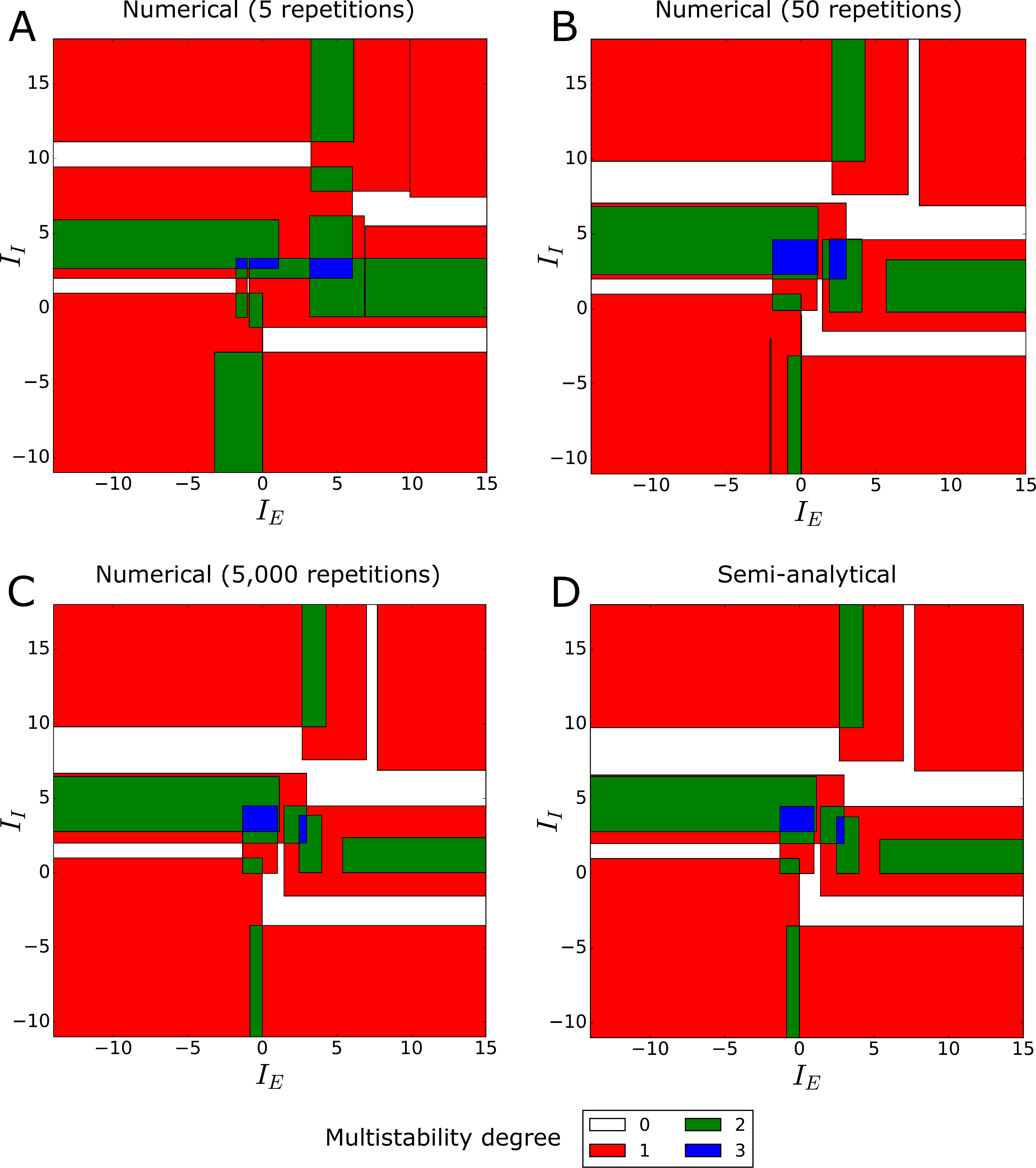}
\par\end{centering}
\caption{\label{fig:mean-multistability-diagram} \small\textbf{ Mean multistability
diagram.} This figure reports the mean multistability diagram of the
small-size network described in Sec.~(\ref{sec:Results}), see Eq.~(\ref{eq:Wigner-semicircle-distribution})
and Tab.~(\ref{tab:Parameters-0}). The diagram shows how the degree
of multistability of the network, namely the number of stationary
solutions, depends on average on the external currents $I_{E}$ and
$I_{I}$. Each color represents a different degree of multistability
$\mathcal{M}$ (e.g., blue = tristability). A) - C) Numerical multistability
diagrams, obtained through Monte Carlo simulations as described in
SubSec.~(\ref{subsec:Numerical-Simulations}), for an increasing
number of network realizations ($5$, $50$ and $5,000$). D) Semi-analytical
multistability diagram, obtained through the techniques described
in SubSec.~(\ref{subsec:Mean-Multistability-Diagram}). Note that,
by increasing the number of network realizations, the numerical multistability
diagram converges to the semi-analytical one.}
\end{figure}

In Fig.~(\ref{fig:mean-multistability-diagram}) we plotted the comparison
between the mean multistability diagram of the network, evaluated
numerically through $5$, $50$ and $5,000$ Monte-Carlo repetitions
(panels A-C), and the same diagram, evaluated semi-analytically through
Eqs.~(\ref{eq:mean_method_1}) or (\ref{eq:mean_method_2}) (panel
D). 
 The figure shows that, by increasing the number of network realizations,
the numerical multistability diagram converges to the semi-analytical
one (compare panels C and D), apart from small numerical errors, that
depend on the integration step in the semi-analytical formulas, and
on the finite number of repetitions in the Monte-Carlo simulations.
The diagrams show a complex pattern of multistability areas in the
$I_{E}-I_{I}$ plane, characterized by multistability degrees $\mathcal{M}=1$
(monostability), $2$ (bistability), and $3$ (tristability). A similar
result was already observed in small binary networks with deterministic
synaptic weights, see \cite{Fasoli2018a,Fasoli2018b}. Moreover, note
that our algorithm detected the presence of white areas, characterized
by multistability degree $\mathcal{M}=0$. In these areas, we did
not observe the formation of stationary firing-rate states, so that
the only possible long-time dynamics for those combinations of stimuli
is represented by neural oscillations. However, generally, oscillations
in the firing-rate states may also co-occur with stationary states
in areas of the $I_{E}-I_{I}$ plane where $\mathcal{M}>0$. The reader
is referred to SubSec.~(\ref{subsec:Future-Directions}) for a discussion
about the possibility to extend our work to the study of neural oscillations.

\begin{figure}
\begin{centering}
\includegraphics[scale=0.26]{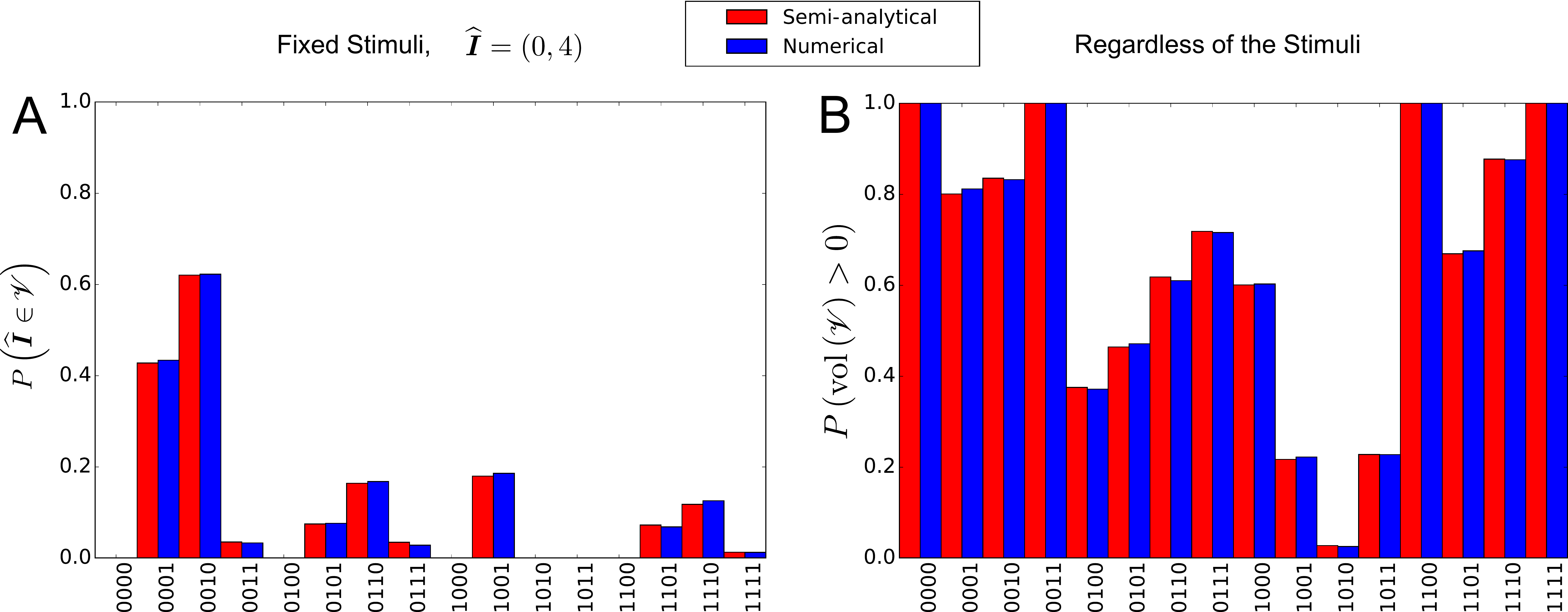}
\par\end{centering}
\caption{\label{fig:occurrence-probability-firing-rate-states} \small\textbf{
Occurrence probability of the firing-rate states.} This figure reports
the occurrence probability of the $2^{N}=16$ firing-rate states of
the network described in SubSec.~(\ref{sec:Results}) (see Eq.~(\ref{eq:Wigner-semicircle-distribution})
and Tab.~(\ref{tab:Parameters-0})), from the state $0000$ to the
state $1111$. A) Occurrence probability of the firing-rate states
for fixed stimuli, obtained for $\widehat{I}_{E}=0$ and $\widehat{I}_{I}=4$.
The red bars represent the occurrence probability calculated semi-analytically
through the method described in SubSec.~(\ref{subsec:Occurrence-Probability-of-the-Stationary-States-for-a-Given-Combination-of-Stimuli}).
The blue bars represent the same probability, evaluated numerically
by a Monte Carlo simulation, as explained in SubSec.~(\ref{subsec:Numerical-Simulations}).
B) Occurrence probability of the firing-rate states, regardless of
the stimuli (red bars calculated according to the approach of SubSec.~(\ref{subsec:Occurrence-Probability-of-the-Stationary-States-Regardless-of-the-Stimuli}),
blue bars computed again through a Monte Carlo simulation). In both
panels, we computed the blue bars over $5,000$ network realizations.
Note that the occurrence probabilities are not normalized (and therefore
they do not represent probability distributions) over the set of $2^{N}$
firing-rate states, see text.}
\end{figure}

In Fig.~(\ref{fig:occurrence-probability-firing-rate-states}) we
plotted the comparison between the semi-analytical and numerical occurrence
probability of the stationary firing-rate states (red and blue bars,
respectively). 
 In panels A and B we plotted, respectively, the probability that
the state $\boldsymbol{\nu}$ is stationary for a fixed combination
of stimuli ($\widehat{I}_{E}=0$ and $\widehat{I}_{I}=4$, red bars
calculated through Eq.~(\ref{eq:probability-for-fixed-stimuli})),
and the probability that $\boldsymbol{\nu}$ is stationary regardless
of the stimuli (red bars calculated through Eq.~(\ref{eq:probability-regardless-of-the-stimuli})).
The figure shows again a very good agreement between semi-analytical
and numerical results.

In particular, panel A shows that, for the network parameters that
we chose (see Tab.~(\ref{tab:Parameters-0})), the states $0000$,
$0100$, $1000$, $1010$, $1011$ and $1100$ are never stationary
for $\widehat{I}_{E}=0$ and $\widehat{I}_{I}=4$. In other words,
in every network realization the rectangles $\mathscr{V}$ corresponding
to these states never contain the point of coordinates $\widehat{\boldsymbol{I}}=\left(\widehat{I}_{E},\widehat{I}_{I}\right)$.
However, panel B shows that, at least for other combinations of the
stimuli, also these firing-rate states can be stationary.

Moreover, panel B shows that the firing-rate states $0000$, $0011$,
$1100$ and $1111$ have unit probability to be observed in the whole
multistability diagram of a single network realization, namely $P\left(\mathrm{vol}\left(\mathscr{V}\right)>0\right)=1$
for these states. This is a consequence of the fact that, for these
states, $\Gamma_{I_{\alpha},u}=\emptyset$ for both $\alpha=E$ and
$\alpha=I$ and for some $u$ (for example, $\Gamma_{I_{E},1}=\Gamma_{I_{I},0}=\emptyset$
for the state $0011$). For this reason, we get $\mathrm{len}\left(\mathcal{V}_{E}\right)=\mathrm{len}\left(\mathcal{V}_{I}\right)=\infty$,
namely $P\left(\mathrm{len}\left(\mathcal{V}_{E}\right)>0\right)=P\left(\mathrm{len}\left(\mathcal{V}_{I}\right)>0\right)=1$,
so that $P\left(\mathrm{vol}\left(\mathscr{V}\right)>0\right)=1$
(see SubSec.~(\ref{subsec:Occurrence-Probability-of-the-Stationary-States-Regardless-of-the-Stimuli})).
On the other hand, for all the other firing-rate states, we obtain
$\Gamma_{I_{\alpha},u}=\emptyset$ only for one value of the index
$\alpha$ (for example, $\Gamma_{I_{E},1}=\emptyset$, $\Gamma_{I_{E},0}=\left\{ 0,1\right\} $,
$\Gamma_{I_{I},1}=\left\{ 3\right\} $ and $\Gamma_{I_{I},0}=\left\{ 2\right\} $
for the state $0001$). Therefore, for these states, typically $P\left(\mathrm{vol}\left(\mathscr{V}\right)>0\right)<1$.

\begin{figure}
\begin{centering}
\includegraphics[scale=0.24]{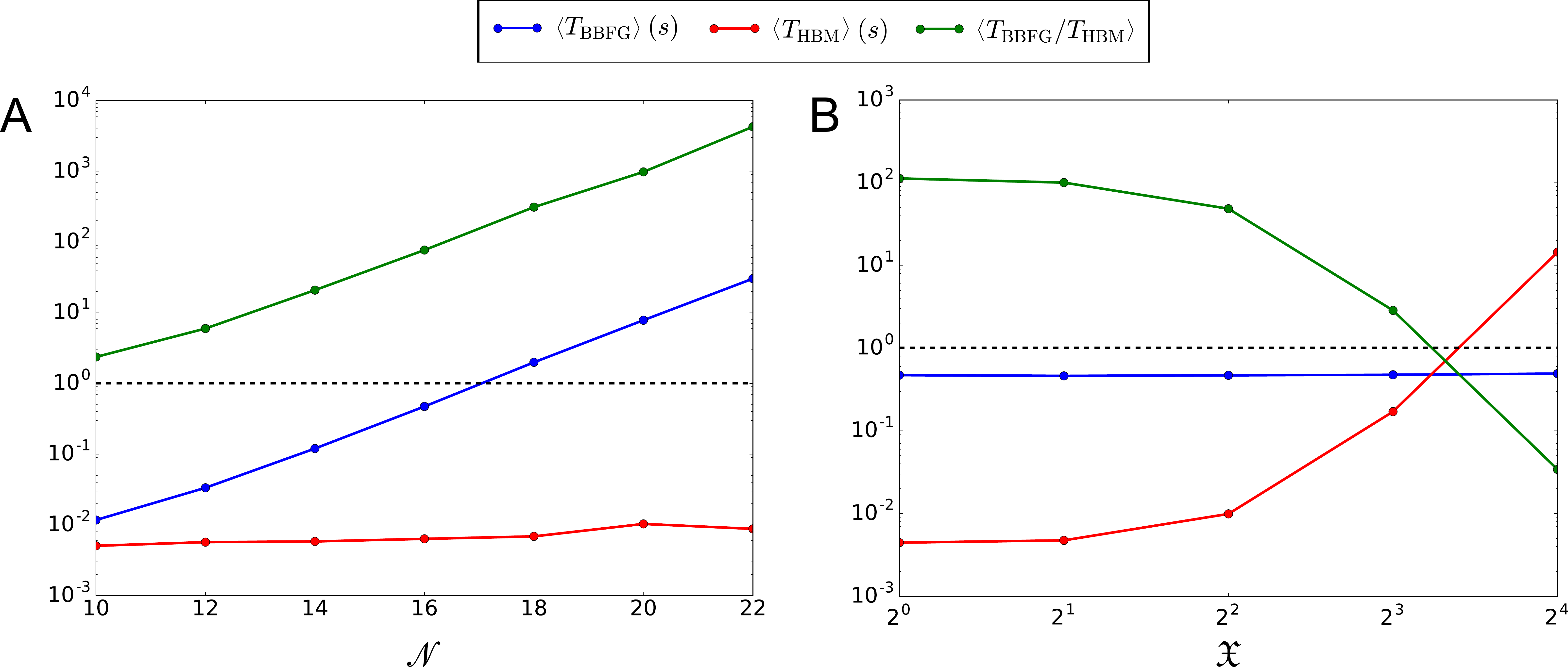}
\par\end{centering}
\caption{\label{fig:speed-test} \small\textbf{ Speed test for the analytical
formula of the permanent.} This figure shows the mean computational
times $\left\langle T_{\mathrm{BBFG}}\right\rangle $ and $\left\langle T_{\mathrm{HBM}}\right\rangle $
(see text) required for calculating the permanent of homogeneous block
matrices $\mathcal{B}$ by means of an Intel\textsuperscript{\textregistered}
Core\texttrademark{} i5-5300U CPU clocked at 2.30GHz with 16 GB RAM.
We chose the entries of the matrices $\mathcal{B}$ to be independent
random numbers $0\leq B_{\lambda,\mu}<0.3$ with two decimal digits,
generated from a uniform probability distribution (see the Python
script ``Permanent.py''), while the remaining parameters of $\mathcal{B}$
are shown in Tab.~(\ref{tab:Parameters-1}). The average times $\left\langle T_{\mathrm{BBFG}}\right\rangle $
and $\left\langle T_{\mathrm{HBM}}\right\rangle $ were calculated
over $100$ repetitions of the matrices. A) Mean computational times
$\left\langle T_{\mathrm{BBFG}}\right\rangle $ (blue line) and $\left\langle T_{\mathrm{HBM}}\right\rangle $
(red line) in seconds, as a function of $\mathscr{N}$. The green
line represents the mean speed gain $\left\langle T_{\mathrm{BBFG}}/T_{\mathrm{HBM}}\right\rangle $
of Eq.~(\ref{eq:permanent-homogeneous-block-matrix}) over the BBFG
algorithm. B) Mean computational times and speed gain as a function
of $\mathfrak{X}$.}
\end{figure}

In Fig.~(\ref{fig:speed-test}) we showed the speed gain of Eq.~(\ref{eq:permanent-homogeneous-block-matrix})
over the BBFG algorithm, achieved during the calculation of the permanent
of homogeneous block matrices that we implemented in the Python script
``Permanent.py''.
The matrices were generated randomly, according to the parameters
in Tab.~(\ref{tab:Parameters-1}).
\begin{table}
\begin{centering}
\textbf{\footnotesize{}}%
\begin{tabular}{|l|l|l|l|}
\hline 
\multicolumn{4}{|c|}{Panel A}\tabularnewline
\hline 
 &  &  & \tabularnewline
$\mathscr{N}=10,\:12,\cdots,\:22$ & $\mathfrak{X}=3$ & $X_{0}=3$ & $Y_{0}=8$\tabularnewline
 & $\mathfrak{Y}=2$ & $X_{1}=5$ & $Y_{1}=\mathscr{N}-8$\tabularnewline
 &  & $X_{2}=\mathscr{N}-8$ & \tabularnewline
 &  &  & \tabularnewline
\hline 
\multicolumn{4}{c}{}\tabularnewline
\multicolumn{4}{c}{}\tabularnewline
\hline 
\multicolumn{4}{|c|}{Panel B}\tabularnewline
\hline 
 &  &  & \tabularnewline
$\mathscr{N}=16$ & $\mathfrak{X}=1,\:2,\:4,\:8,\:16$ & $X_{\lambda}=\frac{\mathscr{N}}{\mathfrak{X}}$ & $Y_{0}=7$\tabularnewline
 & $\mathfrak{Y}=2$ & $\forall\lambda\in\left\{ 0,\cdots,\mathfrak{X}-1\right\} $ & $Y_{1}=9$\tabularnewline
 &  &  & \tabularnewline
\hline 
\end{tabular}
\par\end{centering}{\footnotesize \par}
\caption{\label{tab:Parameters-1} \textbf{Set of Parameters Used for Generating
Fig.~(\ref{fig:speed-test}).}}
\end{table}
 In particular, panel A shows that the mean computational time, that
we called $\left\langle T_{\mathrm{BBFG}}\right\rangle $, required
for calculating the permanent by means of the BBFG algorithm over
several realizations of the matrix, increases exponentially with the
matrix size $\mathscr{N}$. On the contrary, the mean time required
by Eq.~(\ref{eq:permanent-homogeneous-block-matrix}), that we called
$\left\langle T_{\mathrm{HBM}}\right\rangle $, increases very slowly
with $\mathscr{N}$, resulting in a progressive and considerable improvement
of performance over the BBFG algorithm (mean speed gain $\left\langle T_{\mathrm{BBFG}}/T_{\mathrm{HBM}}\right\rangle \gg1$).

Panel B of Fig.~(\ref{fig:speed-test}) shows the limitations of
Eq.~(\ref{eq:permanent-homogeneous-block-matrix}). While $\left\langle T_{\mathrm{BBFG}}\right\rangle $
does not depend on the parameter $\mathfrak{X}$ (namely the number
of neural populations that share the same external stimulus), $\left\langle T_{\mathrm{HBM}}\right\rangle $
strongly decreases with $\mathfrak{X}$, resulting in a progressive
loss of performance of Eq.~(\ref{eq:permanent-homogeneous-block-matrix})
over the BBFG algorithm. This is a consequence of the increasing number
of multinomial coefficients that, according to Eq.~(\ref{eq:permanent-homogeneous-block-matrix}),
must be calculated in order to evaluate the matrix permanent when
$\mathfrak{X}$ is incremented. In more detail, the total number of
multinomial coefficients is:
\begin{spacing}{0.8}
\begin{center}
{\small{}
\[
\sum_{\mu=0}^{\mathfrak{Y}-1}\binom{Y_{\mu}+\mathfrak{X}-1}{\mathfrak{X}-1}\sim\sum_{\mu=0}^{\mathfrak{Y}-1}\frac{\mathfrak{X}^{Y_{\mu}}}{Y_{\mu}!}\left(1+\frac{Y_{\mu}\left(Y_{\mu}+1\right)}{2\mathfrak{X}}+\mathcal{O}\left(\mathfrak{X}^{-2}\right)\right),\quad\mathfrak{X}\leq\sum_{\mu=0}^{\mathfrak{Y}-1}Y_{\mu}=\mathscr{N},
\]
}
\par\end{center}{\small \par}
\end{spacing}

\noindent where the asymptotic expansion holds in the limit $\mathfrak{X}\rightarrow\infty$.
Our analysis shows that generally, in the study of statistically-homogeneous
multi-population networks, Eq.~(\ref{eq:permanent-homogeneous-block-matrix})
should be preferred to the BBFG algorithm when $\mathfrak{X}\ll\mathscr{N}$,
namely when each stimulus is shared by a relatively small number of
populations.

\begin{figure}
\begin{centering}
\includegraphics[scale=0.21]{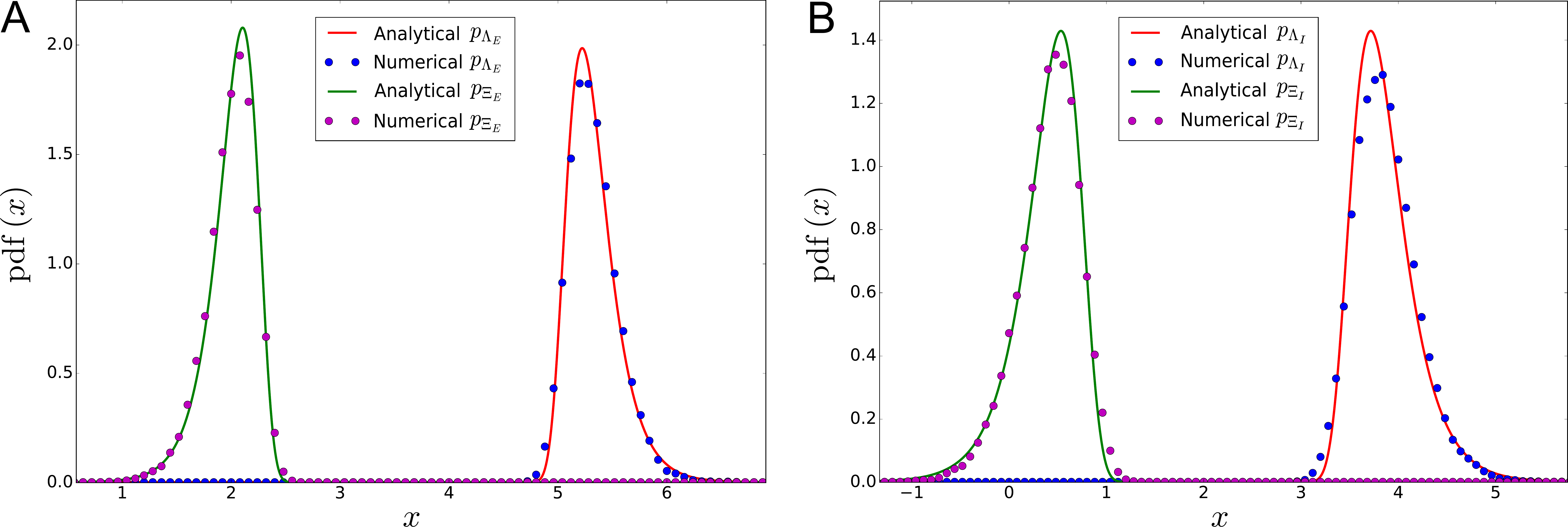}
\par\end{centering}
\caption{\label{fig:large-size-limit} \small\textbf{ Large-size limit of
a statistically-homogeneous two-population network.} This figure shows
the probability distribution of the bifurcation points in a random
network composed of two statistically-homogeneous populations (one
excitatory and one inhibitory) with Laplace-distributed weights (see
Eq.~(\ref{eq:Laplace-distribution})), in the large-size limit. The
parameters of the network are reported in Tab.~(\ref{tab:Parameters-2}).
In particular, note that in this figure we computed the probability
density of the bifurcation points for every firing-rate state $\boldsymbol{\nu}$
that is composed of $\gamma_{E,1}=240$ (respectively $\gamma_{I,1}=80$)
active neurons in the excitatory (respectively inhibitory) population.
The analytical curves (see the green and red solid lines) were obtained
from Eqs.~(\ref{eq:Gumbel-parameters}) and (\ref{eq:Gumbel-pdf-and-cdf}),
while the numerical probability densities (magenta and blue dots)
were calculated over $100,000$ network realizations, as described
in SubSec.~(\ref{subsec:Numerical-Simulations}). A) Analytical and
numerical probability distributions of the the bifurcation points
$\Lambda_{E}$ and $\Xi_{E}$. B) Probability distributions of $\Lambda_{I}$
and $\Xi_{I}$.}
\end{figure}

In Fig.~(\ref{fig:large-size-limit}) we showed examples of the probability
distributions of the bifurcation points in a large random network
composed of two statistically-homogeneous populations (one excitatory
and one inhibitory).
The network parameters that we used are reported in Tab.~(\ref{tab:Parameters-2}).
\begin{table}
\begin{centering}
\textbf{\footnotesize{}}%
\begin{tabular}{|l|l|l|l|l|l|}
\hline 
 &  &  &  &  & \tabularnewline
$N_{E}=640$ & $\gamma_{E,1}=240$ & $\vartheta_{E}=3$ & $\mu_{EE}=11$ & $\sigma_{EE}=0.8$ & $P_{EE}=0.7$\tabularnewline
$N_{I}=160$ & $\gamma_{I,1}=80$ & $\vartheta_{I}=0$ & $\mu_{EI}=-8$ & $\sigma_{EI}=0.6$ & $P_{EI}=0.9$\tabularnewline
 &  &  & $\mu_{IE}=5$ & $\sigma_{IE}=0.7$ & $P_{IE}=1.0$\tabularnewline
 &  &  & $\mu_{II}=-10$ & $\sigma_{II}=0.9$ & $P_{II}=0.8$\tabularnewline
 &  &  &  &  & \tabularnewline
\hline 
\end{tabular}
\par\end{centering}{\footnotesize \par}
\caption{\label{tab:Parameters-2} \textbf{Set of Parameters Used for Generating
Fig.~(\ref{fig:large-size-limit}).}}
\end{table}
 In particular, we supposed that the weights were distributed according
to the following Laplace probability density:
\begin{spacing}{0.8}
\begin{center}
{\small{}
\begin{equation}
p_{W_{i,j}}\left(x\right)=\frac{1}{\sqrt{2s_{\alpha,\beta}}}e^{-\sqrt{\frac{2}{s_{\alpha,\beta}}}\left|x-m_{\alpha,\beta}\right|},\label{eq:Laplace-distribution}
\end{equation}
}
\par\end{center}{\small \par}
\end{spacing}

\noindent $\forall i,\;j$ belonging to populations $\alpha,\;\beta$,
respectively ($m_{\alpha,\beta}$ and $s_{\alpha,\beta}$ are defined
as in Eq.~(\ref{eq:normalization})). Fig.~(\ref{fig:large-size-limit})
shows a good agreement between the analytical formula of the Gumbel
probability density function and numerical simulations, despite slight
differences between analytical and numerical densities can be observed,
as a consequence of the finite size of the network. These differences
disappear in the limit $N\rightarrow\infty$. In particular, Fig.~(\ref{fig:large-size-limit})
shows that the firing-rate states with $240$ active neurons in the
excitatory population, and $80$ active neurons in the inhibitory
one, are very unlikely to be observed in the whole multistability
diagram of the network. For these states, and for our choice of the
network parameters (see Tab.~(\ref{tab:Parameters-2})), the probability
to get $\Xi_{E}>\Lambda_{E}$ and $\Xi_{I}>\Lambda_{I}$ is very small,
as a consequence of the large distance between the peaks of the distributions
of $\Lambda_{\alpha}$ and $\Xi_{\alpha}$. More generally, for a
network composed of an arbitrary number $\mathfrak{P}$ of statistically-homogeneous
populations, we found that the stationary states that are more likely
to occur in the large-size limit are those characterized by homogeneous
intra-population firing rates, namely the stationary states of the
form:
\begin{spacing}{0.8}
\begin{center}
{\small{}
\begin{equation}
\boldsymbol{\nu}=\overset{N_{0}-\mathrm{times}}{\overbrace{\overline{\nu}_{0}\cdots\overline{\nu}_{0}}}\overset{N_{1}-\mathrm{times}}{\overbrace{\overline{\nu}_{1}\cdots\overline{\nu}_{1}}}\cdots\overset{N_{\mathfrak{P}-1}-\mathrm{times}}{\overbrace{\overline{\nu}_{\mathfrak{P}-1}\cdots\overline{\nu}_{\mathfrak{P}-1}}},\quad\overline{\nu}_{\alpha}\in\left\{ 0,1\right\} ,\;\forall\alpha\in\left\{ 0,\cdots,\mathfrak{P}-1\right\} .\label{eq:homogeneous-states}
\end{equation}
}
\par\end{center}{\small \par}
\end{spacing}

\noindent This result proves that, in the large-size limit, the stationary
states of this network can be studied through a dimensional reduction
of the model. In other words, in order to completely characterize
the statistical properties of this network, it suffices to consider
the firing-rate states of the form (\ref{eq:homogeneous-states}),
since the states that present intra-population symmetry breaking are
very unlikely to be observed. The main consequence of this phenomenon
is a tremendous simplification in the mathematical analysis of the
network model, since it reduces the analysis of the $2^{N}$ states
of the network to only $2^{\mathfrak{P}}$ states of the form (\ref{eq:homogeneous-states}).
In turn, this simplification implies a strong reduction of the computational
time of the algorithms, since typically $\mathfrak{P}\ll N$.

\section{Discussion \label{sec:Discussion}}

We studied how the statistical properties of the stationary firing-rate
states of a binary neural network model with quenched disorder depend
on the probability distribution of the synaptic weights and on the
external stimuli. The size of the network is arbitrary and finite,
while the synaptic connections between neurons are assumed to be independent
(not necessarily identically distributed) random variables, with arbitrary
marginal probability distributions. By applying the results derived
in \cite{Vaughan1972,Bapat1989,Bapat1990,Hande1994} for the order
statistics of sets of independent random variables, our assumptions
about the network model allowed us to calculate semi-analytically
the statistical properties of the stationary states and of their bifurcation
points, in terms of the permanent of special matrices.

In particular, in SubSec.~(\ref{subsec:Probability-Distribution-of-the-Bifurcation-Points})
we derived the probability density and the cumulative distribution
functions of the bifurcation points of the model in the stimuli space.
From these distributions, in SubSec.~(\ref{subsec:Mean-Multistability-Diagram})
we derived the mean multistability diagram of the network, namely
the plot of the bifurcation points averaged over network realizations.
Then, in SubSecs.~(\ref{subsec:Occurrence-Probability-of-the-Stationary-States-for-a-Given-Combination-of-Stimuli})
and (\ref{subsec:Occurrence-Probability-of-the-Stationary-States-Regardless-of-the-Stimuli}),
we derived the probability that a given firing-rate state is stationary
for a fixed combination of stimuli, and the probability that a state
is stationary regardless of the stimuli. These results provide a detailed
description of the statistical properties of arbitrary-size networks
with arbitrary connectivity matrix in the stationary regime, and describe
how these properties are affected by variations in the external stimuli.

In SubSec.~(\ref{subsec:The-Special-Case-of-Multi-Population-Networks-Composed-of-Statistically-Homogeneous-Populations})
we specialized to the case of statistically-homogeneous multi-population
networks of arbitrary finite size. For these networks, we found a
compact analytical formula of the permanent, which outperforms of
several orders of magnitude the fastest known algorithm for the calculation
of the permanent, i.e. the Balasubramanian-Bax-Franklin-Glynn algorithm
\cite{Balasubramanian1980,Bax1996,Bax1998,Glynn2010}. Then, in SubSec.~(\ref{subsec:Large-Network-Limit})
we derived asymptotic expressions of the statistical behavior of these
multi-population networks in the large-size limit. In particular,
if the contribution of the autapses to the statistics of the firing
rates can be neglected, we proved that the probability distribution
of the bifurcation point tends to the Gumbel law, and that the statistical
properties of large-size multi-population networks can be studied
through a powerful dimensional reduction.

For the sake of clarity, we implemented our semi-analytical results
for arbitrary-size networks with arbitrary connectivity matrix in
the supplemental Python script ``Statistics.py''. The script performs
also numerical calculations of the probability distributions of the
bifurcation points, of the occurrence probability of the stationary
states and of the mean multistability diagram, through which we validated
our semi-analytical results. To conclude, in the supplemental Python
script ``Permanent.py'', we implemented a comparison between our
analytical formula of the permanent for statistically-homogeneous
multi-population networks, and the Balasubramanian-Bax-Franklin-Glynn
algorithm. This comparison proved the higher performance of our formula
in the specific case of multi-population networks, provided each external
stimulus is shared by a relatively small number of populations.

\subsection{Progress with Respect to Previous Work on Bifurcation Analysis \label{subsec:Progress-with-Respect-to-Previous-Work-on-Bifurcation-Analysis}}

In the study of neural circuits, bifurcation theory has been applied
mostly to networks composed of graded-output units with analog (rather
than discrete) firing rates, see e.g. \cite{Borisyuk1992,Beer1995,Pasemann2002,Haschke2005}.
On the other hand, bifurcation theory of non-smooth dynamical systems
of finite size, including those with discontinuous functions like
the discrete network that we studied in this paper, has recently received
increased attention in the literature. However, the theory has been
developed mostly for continuous-time models \cite{Leine2000,Awrejcewicz2003,Leine2006,Makarenkov2012,Harris2015}
and for piecewise-smooth continuous maps \cite{Parui2002}, while
discontinuous maps have received much less attention, see e.g. \cite{Avrutin2006}.
In \cite{Fasoli2018b} we tackled this problem for finite-size networks
composed of binary neurons with discontinuous activation function
that evolve in discrete-time steps, and we introduced a brute-force
algorithm that performs a semi-analytical bifurcation analysis of
the model with respect to the external stimuli. Specifically, in \cite{Fasoli2018b}
we focused on the study of bifurcations in the case of single network
realizations. In the present paper we extended those results to networks
with quenched disorder, and we introduced methods for performing the
bifurcation analysis of the model over network realizations. While
in \cite{Fasoli2018b} we studied the bifurcations of both the stationary
and oscillatory solutions of the network equations, here we focused
specifically on the bifurcations of the stationary states, while the
study of neural oscillations is discussed in SubSec.~(\ref{subsec:Future-Directions}).

Our work is closely related to the study of spin glasses in the zero-temperature
limit, since a single realization of our network model has deterministic
dynamics. In spin glasses, the physical observables are averaged over
the randomness of the couplings in the large-size limit, by means
of mathematical techniques such as the \textit{replica trick} and
the \textit{cavity method} \cite{Mezard1984,Mezard1986,Mezard2003}.
In our work, we followed a different approach, based on extreme value
theory and order statistics. This allowed us to reduce the mathematical
derivation of the averages and, more generally, of the probability
distributions of the stationary states of arbitrary-size networks,
to the calculation of 1D definite integrals on the real axis.

To our knowledge, bifurcations of neural networks with quenched disorder
were investigated only for fully-connected network models with normally-distributed
weights and graded activation function \cite{Sompolinsky1988,Cessac1995,Faugeras2009,Hermann2012,Cabana2013}.
These studies focused on the thermodynamic limit of the models, preventing
us from making progress in the comprehension of the dynamics of small
networks, such as microcolumns in the primate cortex \cite{Mountcastle1997}
or the nervous system of some invertebrates \cite{Williams1988}.
The neural activity of small networks containing only tens or hundreds
of neurons may show unexpected complexity \cite{Fasoli2018b}. For
this reason, the study of small networks typically requires more advanced
mathematical techniques, because the powerful statistical methods
used to study large networks do not apply to small ones. Contrary
to previous research, in this paper we first focused on the study
of networks of arbitrary size, including small ones. Moreover, unlike
previous work, we considered networks with an arbitrary synaptic connectivity
matrix, which is not necessarily fully connected or normally distributed.
In particular, our work advances the tools available for understanding
small-size neural circuits, by providing a complete (generally semi-analytical)
description of the stationary behavior of Eqs.~(\ref{eq:synchronous-network-equations})
and (\ref{eq:asynchronous-network-equations}). Then, for completeness,
and similarly to \cite{Sompolinsky1988,Cessac1995,Faugeras2009,Hermann2012,Cabana2013},
we studied the large-size limit of multi-population networks composed
of statistically-homogeneous populations. Unlike previous work, which
focused on networks composed of graded-output neurons, our binary-rate
assumption allowed us to derive asymptotic analytical formulas for
the statistics of the stationary states and of the corresponding bifurcation
points, advancing our comprehension of neural networks at macroscopic
spatial scales.

\subsection{Limitations of Our Approach \label{subsec:Limitations-of-Our-Approach}}

A first limitation of the algorithms that we introduced in the supplemental
file ``Statistics.py'' is represented by the network size. Note
that, during the derivation of our semi-analytical formulas in SubSec.~(\ref{subsec:Statistical-Properties-of-the-Network-Model}),
we did not make any assumption about the number of neurons in the
network. As a consequence, our results are exact for networks of arbitrary
size. However, the number of possible firing-rate states in a binary
network grows exponentially with the number of neurons, therefore
in practice our algorithms can be applied only to small-size networks.
The maximum network size that can be studied through our approach
depends on the computational power available.

In order to study the asymptotic statistical properties of large networks,
in this paper we focused on the special case of statistically-homogeneous
networks with arbitrary sparseness and distinct external stimuli to
each neural population. The bifurcation points of these networks obey
the Fisher-Tippett-Gnedenko theorem, in that they correspond to the
extreme values of some independent and identically-distributed random
variables. While the extreme value statistics of a finite number of
(independent and) non-identically distributed samples are known (see
\cite{Bapat1990}), a straightforward generalization of the Fisher-Tippett-Gnedenko
theorem to statistically-heterogeneous samples in the large-size limit
is not available \cite{Kreinovich2017}. For this reason, a second
limitation of our approach is represented by the study of the asymptotic
properties of neural networks whose external stimuli are shared by
two or more populations. In these networks, the extreme value statistics
must be calculated for a set of non-identically distributed random
variables, see our discussion at the end of SubSec.~(\ref{subsec:Large-Network-Limit}).
Therefore the complete characterization of these networks still represents
an open problem.

Similarly to \cite{Sompolinsky1988,Cessac1995,Faugeras2009,Hermann2012,Cabana2013},
a third limitation of our work is represented by the assumption of
statistical independence of the synaptic connections. The calculation
of order statistics for dependent random variables represents another
open problem in the literature, which prevents the extension of our
results to neural networks with correlated synaptic connections. In
computational neuroscience, the dynamical and statistical properties
of this special class of neural networks are still poorly understood.
A notable exception is represented by \cite{Faugeras2015}, which
provides a theoretical study of a graded-rate network with correlated
normally-distributed weights in the thermodynamic limit.

\subsection{Future Directions \label{subsec:Future-Directions}}

Stationary states represent only a subset of the dynamic repertoire
of a binary network model. In future work, we will investigate the
possibility to extend our results to neural oscillations. In particular
note that, according to \cite{Fasoli2018b}, the bifurcation points
at which an existing neural oscillation disappears, or the formation
of a new oscillation is observed, correspond to the minima of minima
or to the maxima of maxima of sets of random variables. Provided the
arguments of the functions $\mathrm{min}\left(\cdot\right)$ and $\mathrm{max}\left(\cdot\right)$
are independent, it follows that, in principle, the statistics of
neural oscillations could be studied (semi-)analytically by applying
extreme value theory twice.

\section*{Acknowledgments}

This research was in part supported by the Flag-Era JTC Human Brain
Project (SLOW-DYN).

\noindent The funders had no role in study design, data collection
and analysis, decision to publish, interpretation of results, or preparation
of the manuscript.

\bibliographystyle{plain}
\bibliography{Bibliography}

\end{document}